\documentclass[twocolumn]{aastex63}

\usepackage{newtxtext}

\usepackage[T1]{fontenc}
\usepackage{ae,aecompl}
\usepackage{color,comment}
\usepackage[normalem]{ulem}

\newcommand{\ben}[1]{}
\newcommand{\aaron}[1]{}
\newcommand{\aarono}[1]{}

\usepackage{graphicx}
\usepackage{amsmath}
\usepackage{amssymb}
\usepackage{float}
\usepackage{wasysym}    

\makeatletter 
  \patchcmd{\NAT@citex}
    {\@citea\NAT@hyper@{%
      \NAT@nmfmt{\NAT@nm}%
      \hyper@natlinkbreak{\NAT@aysep\NAT@spacechar}{\@citeb\@extra@b@citeb}%
      \NAT@date}}
    {\@citea\NAT@nmfmt{\NAT@nm}%
    \NAT@aysep\NAT@spacechar\NAT@hyper@{\NAT@date}}{}{}

  \patchcmd{\NAT@citex}
    {\@citea\NAT@hyper@{%
      \NAT@nmfmt{\NAT@nm}%
      \hyper@natlinkbreak{\NAT@spacechar\NAT@@open\if*#1*\else#1\NAT@spacechar\fi}%
        {\@citeb\@extra@b@citeb}%
      \NAT@date}}
    {\@citea\NAT@nmfmt{\NAT@nm}%
    \NAT@spacechar\NAT@@open\if*#1*\else#1\NAT@spacechar\fi\NAT@hyper@{\NAT@date}}
    {}{}
\makeatother

\newcommand\Msun{\text{M}_{\astrosun}} 
\newcommand\HI{{H\,\textsc{i}}}
\newcommand\HII{{H\,\textsc{ii}}}

\begin{document}

\title{The Origin and Evolution of Lyman-$\alpha$ Blobs in Cosmological Galaxy Formation Simulations}
\shorttitle{The Origin and Evolution of LABs}
\shortauthors{Kimock et al.}

\author[0000-0002-3160-7679]{Benjamin Kimock}
\affiliation{Department of Astronomy, University of Florida, 211 Bryant Space Sciences Center, Gainesville, Florida, 32611, USA}
\author[0000-0002-7064-4309]{Desika Narayanan}
\affiliation{Department of Astronomy, University of Florida, 211 Bryant Space Sciences Center, Gainesville, Florida, 32611, USA}
\affiliation{University of Florida Informatics Institute, 432 Newell Drive, CISE Bldg E251, Gainesville, FL 32611}
\affiliation{Cosmic Dawn Center at the Niels Bohr Institute, University of Copenhagen and DTU-Space, Technical University o\
f Denmark}
\author[0000-0002-2838-9033]{Aaron Smith}
\altaffiliation{NHFP Einstein Fellow}
\affiliation{Department of Physics, Massachusetts Institute of Technology, Cambridge, MA 02139, USA
}
\author[0000-0001-8091-2349]{Xiangcheng Ma}
\affiliation{Department of Astronomy and Theoretical Astrophysics Center, University of California Berkeley, Berkeley, CA 94720, USA; TAPIR, MC 350-17, California Institute of Technology, Pasadena, CA 91125, USA}
\author[0000-0002-1109-1919]{Robert Feldmann}
\affiliation{Institute for Computational Science, University of Zurich, Zurich CH-8057, Switzerland, UK}
\author[0000-0001-5769-4945]{Daniel Angl\'es-Alc\'azar}
\affiliation{Center for Computational Astrophysics, Flatiron Institute, 162 Fifth Avenue, New York, NY 10010, USA}
\affiliation{Department of Physics, University of Connecticut, 196 Auditorium Road, U-3046, Storrs, CT 06269-3046, USA}
\author{Volker Bromm}
\affiliation{Department of Astronomy, The University of Texas at Austin, Austin, TX 78712, USA}
\author[0000-0003-2842-9434]{Romeel Dave}
\affiliation{Institute for Astronomy, Royal Observatory, Univ. of Edinburgh, Edinburgh EH9 3HJ, UK}
\affiliation{University of the Western Cape, Bellville, Cape Town 7535, South Africa}
\affiliation{South African Astronomical Observatories, Observatory, Cape Town 7925, South Africa}
\author[0000-0003-4964-4635]{James E. Geach}
\affiliation{Centre for Astrophysics Research, School of Physics, Astronomy and Mathematics, University of Hertfordshire, College Lane, Hatfield AL10 9AB, UK}
\author[0000-0003-3729-1684]{Philip Hopkins}
\affiliation{TAPIR, Mailcode 350-17, California Institute of Technology, Pasadena, CA 91125, USA}
\author[0000-0002-1666-7067]{Du\u{s}an Kere\u{s}}
\affiliation{Department of Physics, Center for Astrophysics and Space Science, University of California at San Diego, 9500 Gilman Drive, La Jolla, CA 92093, USA}

\correspondingauthor{Benjamin Kimock}
\email{kimockb@gmail.com}

\begin{abstract}
    High-redshift Lyman-$\alpha$ blobs (LABs) are an enigmatic class of objects that have been the subject of numerous observational and theoretical investigations.
    It is of particular interest to determine the dominant power sources for the copious luminosity, as direct emission from \HII\ regions, cooling gas, and fluorescence due to the presence of active galactic nuclei (AGN) can all contribute significantly.
    In this paper, we present the first theoretical model to consider all of these physical processes in an attempt to develop an evolutionary model for the origin of high-$z$ LABs.
    This is achieved by combining a series of high-resolution cosmological zoom-in simulations with ionization and Lyman~$\alpha$ (Ly$\alpha$) radiative transfer models.
    We find that massive galaxies display a range of Ly$\alpha$ luminosities and spatial extents (which strongly depend on the limiting surface brightness used) over the course of their lives, though regularly exhibit luminosities and sizes consistent with observed LABs.
    The model LABs are typically powered from a combination of recombination in star-forming galaxies, as well as cooling emission from gas associated with accretion. \aaron{I suggest reporting numbers here (the abstract generally) so people can already get a feel for the results. Okay, possibly Gini $G \approx 0.9$ to distinguish AGN, but the abstract is fine as is.}\ben{We are concerned the abstract is long already and there isn't a number we can put here that's easy to understand at a glance. If anyone has a metric in mind I'll entertain it.}
    When AGN are included in the model, the fluorescence caused by AGN-driven ionization can be a significant contributor to the total Ly$\alpha$ luminosity as well.
    We propose that the presence of an AGN may be predicted from the Gini coefficient of the blob's surface brightness.
    Within our modeled mass range, there are no obvious threshold physical properties that predict appearance of LABs, and only weak correlations of the luminosity with the physical properties of the host galaxy.
    This is because the emergent Ly$\alpha$ luminosity from a system is a complex function of the gas temperature, ionization state, and Ly$\alpha$ escape fraction.
\end{abstract}


\section{Introduction}
\label{sec:introduction}

Lyman-$\alpha$ ``blobs'' (LABs) are an enigmatic class of objects first discovered roughly two decades ago \citep{Fynbo1999,Steidel2000}, and are characterized by their copious Lyman~$\alpha$ (Ly$\alpha$) luminosities and large spatial extents.
While there is no consensus for the definition of a LAB, the majority of blobs have luminosities in excess of $\sim 10^{43}$\,erg/s, and spatial extents that greater than $\sim50$ kpc in radius.
We will discuss these definitions in more detail shortly.

Since their discovery, the dominant source of power in these objects has been under debate.
Broadly, there are two major sources: (\textit{i}) in situ emission from \HII\ regions surrounding actively star-forming regions in the central galaxy or satellites \citep[e.g.][]{Geach2016} that is subsequently scattered in the circumgalactic gas, and (\textit{ii}) direct radiation from extended gas in the halo.
Emission from the latter can be due to either cooling radiation from a collisionally excited circumgalactic medium \citep[e.g.][]{Katz1996,Haiman2000,Fardal2001}, or photoionized gas originating from the meta-galactic UV background (UVB) and starbursts or active galactic nuclei (AGN) driving Ly$\alpha$ production via \HII\ recombination \citep{Kollmeier2010,Gronke2017}.

Claims of LABs powered by cooling accretion from the intergalactic medium (IGM) are often observationally justified by the detection of LABs without any observable AGN.
For example, \citet{Smith2007} observed a blob at $z=2.83$ for which they are able to rule out AGN based on non-detections of highly ionized lines.
Similarly, \citet{Smith2007} rule out direct emission from \HII\ regions based on a relatively low derived SFR from the UV continuum of  $\sim 25\,\Msun\,\text{yr}^{-1}$.
 \citet{Scarlata2009} identified a LAB that is associated with two galaxies, and present spectroscopic evidence against emission driven by star formation or AGN, as they do not detect C\,\textsc{iv} or N\,\textsc{v} lines.
\citet{Nilsson2006} also argue against the presence of AGN or super-winds based on the lack of a continuum counterpart detection in a $z=3.16$ blob \citep[athough this is debated, see e.g.][]{Prescott2015}.

At the same time, other studies argue heavily for AGN-driven fluourescence.
For example, some LABs are radio loud \citep{Miley2008}, with correlated Ly$\alpha$ and radio extents.
This correlation implies that a central AGN may be powering the extended Ly$\alpha$ \citep{vanOjik1997}.
Indeed, the heavily-studied LAB~1 appears to be powered by a hidden quasar \citep{Overzier2013}, based on observations of [O\,\textsc{iii}], and it is argued that AGN may power the most luminous LABs.
\citet{Geach2009} report X-ray observations of LABs, finding an AGN fraction of $17^{+12}_{-7}\%$, but with all (5 of 29) detections they find heavy obscuration and suggest that there may be heavily obscured AGN in many LABs.

The tendency of LABs to appear in over-dense environments \citep{Matsuda2009,Matsuda2011,Prescott2008} suggests that the power source may relate to elevated star formation rates typically associated with the formation of massive galaxies \citep[e.g.][]{Matsuda2007,Kubo2013,Hine2016,Alexander2016}.
However, care should be taken in assessing the role of star formation in powering LABs, since the signatures of elevated star formation and AGN activity can be nearly identical \citep{Webb2009}.

LABs may also be powered indirectly by AGN or star formation through galaxy-scale winds \citep{Wilman2005}.
Based on detections of bubbles in LAB~1, \citet{Matsuda2004} deduce a SFR $\sim 600\,\Msun\ \text{yr}^{-1}$, which is in agreement with submillimeter observations \citep{Chapman2001}.
\citet{Matsuda2007} also argue for the possibility of extended starbursts or winds on the basis of correlated submilimiter and Ly$\alpha$ emission in LAB~1.
Additionally, \citet{Ohyama2003} interpret the double-peaked Ly$\alpha$ spectrum, particularly the decrease in the velocity separation of the two peaks with distance from the center of LAB~1 as evidence for wind-driven Ly$\alpha$.

The last two decades of observations have brought little consensus on the dominant source(s) of emission in Ly$\alpha$ blobs, or even whether a single physical process dominates.
Indeed, some authors think that LABs may be powered by a variety of mechanisms \citep{Scarlata2009,Webb2009,Nilsson2006,Prescott2009,Ao2015}.
Additionally, the physics of Ly$\alpha$ escape from high-redshift galaxies remains an open problem in connection with LABs, though it is deeply coupled to the emission thereof \citep{Smith2019}.
This leaves these massive objects largely unexplained in spite of their relevance to massive galaxy formation and reionization, as we do not understand what if anything the Ly$\alpha$ traces.

\subsection{The Definition of a Ly\texorpdfstring{$\alpha$}{a} Blob}
\label{section:blob_definition}
There is no consensus definition of a Ly$\alpha$ blob in the
literature.  We present in Table~\ref{table:labs} a summary of recent
papers aimed at observationally characterizing LABs, and quote
their measurements of a few observed properties that could potentially
be used to distinguish this class of objects from Lyman-alpha
emitters (LAEs).  As is evident, there is no clear luminosity threshold for a
LAB definition.  Observations find luminosities ranging nearly two
orders of magnitude ($2\times10^{42}\,\text{erg/s} < L_{\text{Ly}\alpha} < 2.1\times10^{44}\,\text{erg/s}$).  Similarly, there is no clear
size definition.  Quoted diameters range from $30$--$200$\,kpc, though
the interpretation of this physical constraint is muddied by the fact that observations have a wide range of surface brightness sensitivities that range by over an order of magnitude in the literature. 
Beyond this, the dispersion in this limiting surface brightness along with the
amorphous morphology of LABs makes such size measurements
difficult to interpret.  To further complicate matters, recent work by
\citet{Wisotzki2018} has shown that with sufficient sensitivity nearly
the whole sky is covered by Ly$\alpha$. Indeed, as we will
demonstrate in this work, the area enclosed by a Ly$\alpha$ blob is a
strong function of the limiting surface brightness.

Going forward in this paper, we will adopt a notional threshold
luminosity for a blob definition of $L_{\text{Ly}\alpha} > 10^{43}$\,erg/s,
with no size threshold.  This said, we will explore the impact of
modifying this on our results.

\subsection{Theoretical Efforts to Date}

\citet{Furlanetto2005}, \citet{Laursen2007}, \citet{Cen2013}, \citet{Geach2016}, and \citet{Gronke2017} have studied the contribution of star formation on the formation of LABs, and all concluded this source of Ly$\alpha$ can (or in the case of \citet{Cen2013} \emph{must}) power blobs.
Additionally, \citet{Cen2013} were able to reproduce an observed LAB luminosity-size relation.
However, some of the previous work on the contribution of star formation relies on a simplified SFR--$L_{\text{Ly}\alpha}$ conversion based on the expected luminosity from case~B recombination.

There has also been extensive study of the contribution of Ly$\alpha$ emission due to collisionally excited neutral hydrogen \citep{Rosdahl2012, Fardal2001, Goerdt2010, Haiman2000, Faucher-Giguere2010}, or specifically the incoming streams of cooling IGM that are observed in some simulations at high redshift.
These works are able to reproduce the requisite Ly$\alpha$ luminosity to power a LAB, but sometimes have difficulty with the particular appearance of LABs in surface brightness maps.

Other authors have studied the effect of fluorescence from an external ionizing field such as a nearby or internal quasar \citep{Haiman2001} or the cosmological UV background or winds \citep{Furlanetto2005, Mas-Ribas2016}.
These authors find that an external ionizing radiation field can produce extended Ly$\alpha$ emission, but not quite at the surface brightnesses to produce a blob on its own.

Missing, to date, is a comprehensive model that considers all of these physical processes simultaneously.  In this paper, we attempt to provide just that.   We present a model for the formation and evolution of Ly$\alpha$ blobs by combining high-resolution cosmological zoom-in simulations with ionization radiative transfer and Ly$\alpha$ radiative transfer.  We consider the physics of emission from ionized gas surrounding massive stars, cooling, and fluourescence induced from an external ionizing field (including AGN).  In \S~\ref{sec:methods}, we detail our numerical methodology. In \S~\ref{section:evolution}, we describe the evolution of the Ly$\alpha$ luminosity from massive galaxies at high-redshift.  We follow this in \S~\ref{sec:origins} with an investigation into the dominant power sources of Ly$\alpha$ photons in massive galaxies.  We investigate the role of AGN in \S~\ref{sec:agn}, and discuss the broader physical properties of our model LABs in \S~\ref{section:general_physical_properties}.  We provide discussion in \S~\ref{sec:discussion} and conclude in \S~\ref{sec:conclusions}.

\begin{table*}
\caption{
    An overview of Ly$\alpha$ properties for a sample of known LABs which we compare our simulations to.
}
\centering
\begin{tabular}{ | l | c | c | c | c | }
\hline
     & & $L_{\text{Ly}\alpha}$ & $\Sigma_\text{lim}$ & Area \\
    Publication & $z$ & $(10^{43}\,\text{erg\ s}^{-1})$ & $(10^{-18}\,\text{erg\ s}^{-1}\,\text{cm}^{-2}\,\text{arcsec}^{-2})$ & (arcsec$^{2}$) \\
\hline

\citet{Matsuda2004} & 3.1 & $0.9$--$11.0$ & $2.2$ & 16--222\\
\hline

\citet{Nilsson2006} & 3.16 & $1.0$ & $3.7$ & 47 \\
\hline

\citet{Smith2007} & 2.83 & $21.0$ & -- & 110\\
\hline

\citet{Ouchi2009} & 6.595 & $3.9$ & $1.63$ & 7 \\
\hline

\citet{Yang2009} & 2.3 & $1.6$--$5.3$ & $2.47$ & 25 \\
\hline

\citet{Matsuda2011} & 3.09 & $0.8$--$20.4$ & $1.4$ & 28--181\\
\hline

\citet{Steidel2011} & 2.65 & $6.57$ & $\sim1$ & 30 \\
\hline

\citet{Barger2012} & 0.977 & $0.72$ & -- & 500 \\
\hline

\citet{Prescott2013} & 1.7--2.7 & $0.26$--$1.9$ & $0.933$ & 5.9--104 \\
\hline

\citet{Caminha2016} & 3.118 & $0.19$ & -- & 14 \\
\hline

\citet{Badescu2017} & 2.3 & $0.9$--$1.3$ & $2.1$ & 10--12 \\ 
\hline

\citet{North2017} & 3.08 & $2.2$ & $7.5$ & 7--12\\
\hline

\citet{Shibuya2017} & 5.7--6.6 & $1.26$--$7.94$ & $10.0$--$21.0$ & 2--3\\
\hline

\end{tabular}
\label{table:labs}
\end{table*}

\section{Methods}
\label{sec:methods}
\subsection{Overview}
\label{sec:pipeline}
Our overall goal is to extract Ly$\alpha$ observables from cosmological zoom-in simulations of massive galaxies in evolution in post-processing.
To do this, we construct a pipeline in which we smooth the particle data onto an octree grid on which we perform ionizing radiative transfer, to determine the ionization state of the gas in the halo.
We then perform Ly$\alpha$ radiative transfer calculations in order to compute both the intrinsic Ly$\alpha$ luminosity (considering both recombination and collisional processes), as well as the escape fraction from the region.
In what follows, we go into substantially more detail about each of these numerical techniques.
The reader who is less interested in our numerical implementation may skip the remainder of this section without loss of continuity, though we encourage them to peruse Tables~\ref{table:sims2} and~\ref{table:sims5} for an overview of the physical properties of our model galaxies.

\subsection{Cosmological Hydrodynamic Zoom Galaxy Formation Simulations}
The galaxy formation simulations studied here were presented in \citet{Anglesalcazar2017b}, which are part of the Feedback In Realistic Environments (FIRE) project\footnote{See the FIRE project web site at: \href{http://fire.northwestern.edu}{http://fire.northwestern.edu}.}.
These simulations employ the FIRE-2 suite of physics \citep{Hopkins2018}, and their initial conditions are derived from the MassiveFIRE suite \citep{Feldmann2016}.
These physics modules are fully described in \citet{Hopkins2018}, and we point the reader to this work, summarizing the salient details.

The initial conditions for the MassiveFIRE simulations are generated with \textsc{music} \citep{Hahn2011} for a $(100\,\text{Mpc}/h)^3$ box.
We first run an initial low-resolution simulation, from which we selected particular halos for re-simulation at much higher resolution.
For these halos, the region encompassing the high-resolution particles was selected with a convex hull filter selecting all particles within $3$ virial radii of the halo at $z=1.$
These particles were then split to obtain higher mass resolution, and the entire simulation was re-run with a final mass resolution of the high-resolution particles of $m_\text{DM} = 1.7 \times 10^5\,\Msun$ and $m_\text{baryon} = 3.3 \times 10^4\,\Msun$ for dark matter and baryons, respectively.

The simulations themselves are run with \textsc{gizmo} \citep{Hopkins2015} with the hydrodynamics run in Meshless Finite Mass (MFM) mode.
These simulations include star formation in dense and self-gravitating gas \citep*{Hopkins2013}, and stellar feedback channels that include radiation pressure, photoionization, photoelectric heating, O-star and Asymptotic Giant Branch (AGB) driven stellar winds, and Type I and II Supernovae.
Supermassive black holes are included in the simulations, but followed passively, meaning that feedback from an active galactic nuclei (AGN) is not included.
This said, black holes accreted following the torque-limited accretion model of \citet{Anglesalcazar2017a,Anglesalcazar2017b}.
In this paper, we examine $4$ massive halos, whose physical properties are described at $z=2$ and $z=5$ in Tables~\ref{table:sims2} and ~\ref{table:sims5} respectively.
These halos have the same initial conditions as those monikered ``A1'', ``A2'', ``A4'', and ``A8'' from \citet{Feldmann2016}, though are different from the original halos in that they are run with FIRE-2 physics (the original halos presented in Feldmann et al. are run with FIRE-1 physics \citep{Hopkins2014} and did not include SMBHs).
See \citet{Cochrane2019} and \citet{Wellons2019} for detailed studies of the spatially-resolved dust continuum emission in the central galaxies of these halos as well as their kinematic and structural properties.

\begin{table*}
\caption{Physical properties for the simulations we use at $z=5$, within our $(150\,\text{kpc})^3$ box.}
\centering
\begin{tabular}{ | l | r | r | r | r | }
\hline
    Name & $M_\text{DM}$ & $R_\text{vir}$\footnote{Calculated using \citet{Bryan1997}} & $M_\text{star}$ & SFR \\
     & ($\Msun$)  & (kpc)         & ($\Msun$)    & ($\Msun / \text{yr}^{-1}$)\\ \hline
A1 & $9.72\times 10^{11}$ & $5.60 \times 10^1$  & $2.07\ \times 10^{10}$ & $3.11\times10^1$ \\ \hline
A2 & $3.98\times 10^{11}$ & $4.20 \times 10^1$  & $3.81 \times 10^{9}$ & $2.65\times10^1$ \\ \hline
A4  & $2.98\times 10^{11}$ & $3.81 \times 10^1$  & $1.32 \times 10^{9}$ & $4.70\times10^0$ \\ \hline
A8 & $3.04\times 10^{11}$ & $3.84 \times 10^1$ & $9.38 \times 10^{8}$ & $8.18\times10^0$ \\
\hline
\end{tabular}
\label{table:sims2}
\end{table*}

\begin{table*}
\caption{Physical properties for the simulations we use at $z=2$, within our $(150\,\text{kpc})^3$ box.}
\centering
\begin{tabular}{ | l | r | r | r | r | }
\hline
    Name & $M_\text{DM}$ & $R_\text{vir}$\footnote{Calculated using \citet{Bryan1997}} & $M_\text{star}$ & SFR \\
     & ($\Msun$)  & (kpc)         & ($\Msun$)    & ($\Msun / \text{yr}^{-1}$)\\ \hline
A1 & $1.64\times 10^{12}$ & $1.20 \times 10^2$  & $1.78 \times 10^{11}$ & $6.55\times10^1$ \\ \hline
A2 & $2.00\times 10^{12}$ & $1.29 \times 10^2$  & $2.98 \times 10^{11}$ & $1.68\times10^2$ \\ \hline
A4  & $1.63\times 10^{12}$ & $1.19 \times 10^2$  & $1.41 \times 10^{11}$ & $7.15\times10^1$ \\ \hline
A8 & $1.92\times 10^{12}$ & $1.26 \times 10^2$ & $8.06 \times 10 ^{10}$ & $8.79\times10^1$ \\
\hline
\end{tabular}
\label{table:sims5}
\end{table*}

\begin{figure}
    \centering
    \includegraphics[width=\columnwidth]{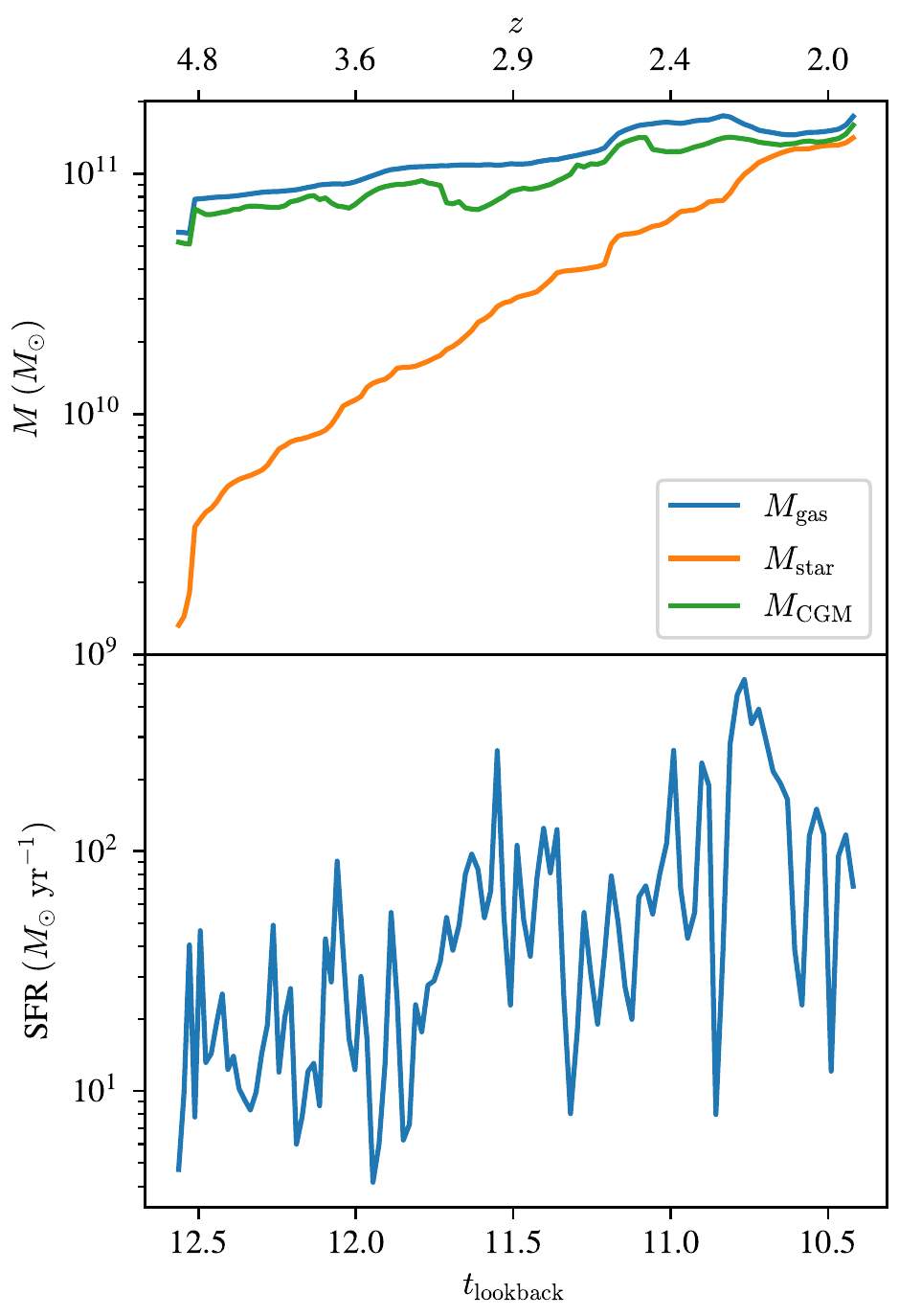}
    \caption{Evolution of physical properties for a sample halo (A4).
    We show (from top to bottom) the total gas mass, stellar mass, star formation rate, and circumgalactic medium (CGM) gas, defined as all gas in the box not associated with the central galaxy.
    The physical properties are computed over the $(150\,\text{kpc})^3$ box employed for our radiative transfer calculations.}
    \label{fig:properties_redshift}
\end{figure}

\subsection{Computing the Ionization State of the Gas}
\label{sec:lycrt}
The first post-processing step in the cosmological galaxy formation
simulations is to compute the ionization state of the gas.  To do
this, we employ \textsc{lycrt} \citep{Ma2015, Ma2020}.

\textsc{Lycrt} is a Monte Carlo radiative transfer code that iteratively
computes the gas ionization state by emitting rays from both stars and
a UV background. The radiative transfer is performed on an octree
grid, containing the smoothed information from the particle data. These
rays are subject to absorption by \HI\ along with dust scattering and
absorption, using a constant dust-to-metal ratio of 0.4 below $10^6$ K, no dust above $10^6$ K, and the SMC grain size from \citet{Weingartner2001}.
The passage of these rays through octree cells is used to compute the ionizing UV field.

The ionization state of the gas is updated based on the UV field after
each iteration, and includes collisions with electrons when solving
for the thermal state \citep{Fumagalli2011}\footnote{Since our ionization solver does not update the gas temperature the ionization state is self-inconsistent: the gas surrounding massive stars may be underheated. However, this innacuracy likely induces only a small change to the total Ly$\alpha$ emissivity as the recombination coefficient varies weakly with temperature.}.
We assume a redshift-dependent UV background with intensity as determined by the model of
\citet{Faucher-Giguere2009}.  The stellar spectral energy distributions (SEDs) are generated with
Binary Population and Spectral Synthesis (\textsc{bpass}) spectral
libraries that include the effect of binary stars
\citep{Eldridge2008}.

Later in this work we will use this ionization state calculation to add an approximate model for the efffects of AGN, wherein we effectively treat the AGN as a very bright star particle.
A complete description of how we derive the ionizing intensity for each AGN is included in \S~\ref{sec:agn}, along with a discussion of their effects.

\subsection{Ly\texorpdfstring{$\alpha$}{a} Emission and Radiative Transfer}
\label{sec:colt}
 
The Ly$\alpha$ Monte Carlo radiative transfer calculations are performed using the Cosmic Ly$\alpha$ transfer Code \textsc{colt} \citep{Smith2015}. \textsc{colt} models the  emission of Ly$\alpha$ photons due to hydrogen recombination and radiative de-excitation of collisionally excited hydrogen, and accounts for scattering due to neutral hydrogen, and scattering and absorption due to dust.

To do this, \textsc{colt} generates Monte Carlo photon packets in octree
cells with probability proportional to the Ly$\alpha$ luminosity of
each cell. We provide a more detailed discussion of the recombination and collisional emission processes in \S~\ref{sec:physicalconcepts}. \footnote{In addition to the underheating issues described in Footnote 2, the {\sc FIRE} simulations include only a crude treatment of self-shielding in the gas, which can drive temperatures in the {\it opposite} direction, i.e. toward overheating.  This can impact the emergent Ly$\alpha$ luminosity substantially due to the strong dependence of the collisional rate coefficients on temperature \citep[e.g.][]{Faucher-Giguere2010}. In Appendix ~\ref{app:half_temperature}, we investigate the impact of these temperature innaccuracies in detail.}.
Photons are sampled uniformly over the
unit sphere, and the sub-grid positions are randomly distributed
within cell volumes. The transport of photon packets follows the usual
Monte Carlo scheme, with the local Ly$\alpha$ absorption coefficient
given by
\begin{equation}
    \label{eq:kalpaha}
    k_\alpha = n_\text{\HI\,} \sigma_\alpha(\nu) \, ,
\end{equation}
where $\sigma_\alpha(\nu)$ is the Ly$\alpha$ cross-section, which at
line center is $5.9 \times 10^{-14}\,[T / (10^4\,\text{K})]^{-1/2}\,\text{cm}^2$. Following
\citet{Laursen2009}, we assume SMC-like dust properties with an
effective cross-section per hydrogen atom of
$3.95\times10^{-22}\,\text{cm}^{2}$, and fiducial albedo of 0.32.
We do not alter the mechanisms for scattering, but do make an adjustment to how dust absorption is accounted for.
Instead of drawing a random variable to determine if a photon packet is absorbed by an interaction with dust (as done by \citet{Laursen2007} and \citet{Smith2015}) we treat dust absorption as a continuous process.
Each photon packet has a weight that is attenuated at each scattering by the amount of dust it traversed since its last scattering.
Therefore, we can compute the escape fraction from the simulation as the sum of the weights of all photon packets upon escaping the simulation.

Beyond the standard \textsc{colt} algorithms, we have developed a number
of performance-enhancing features.  For example, to avoid following
extremely low weight photon paths we introduce a conservative
traversed optical depth threshold, after which we discard photons from
the simulation. We verified that each algorithmic change does not
alter the final luminosity or surface brightness of our model
galaxies.  Similarly, we have introduced updated communication modules
into \textsc{colt} that distribute photon packets individually with
asynchronous message passing interface (MPI) communication patterns,
as opposed to batches of photons. We find this allows for a nearly
$30\%$ speedup due to the large variance in computation time per
photon packet within photon batches, i.e. this minimizes the time
individual nodes sit idle.

\section{Lyman-\texorpdfstring{$\alpha$}{a} Histories of Massive Halos}
\label{section:evolution}
\subsection{Evolution of Massive Galaxies}

To help orient the reader, in Figure~\ref{fig:properties_redshift}, we
first plot the evolution of one of our model simulations, halo A4.  We
pick this model as we will use it throughout the paper as a fiducial
galaxy to examine, though note that the results presented in this
paper are generic to all of our model halos.  The physical properties
presented in Figure~\ref{fig:properties_redshift} are not of the
central galaxy or halo, but rather the $(150\,\text{kpc})^3$ box over
which we will perform our radiative transfer simulations.

\begin{figure*}
  \centering
  \includegraphics[width=2.1\columnwidth]{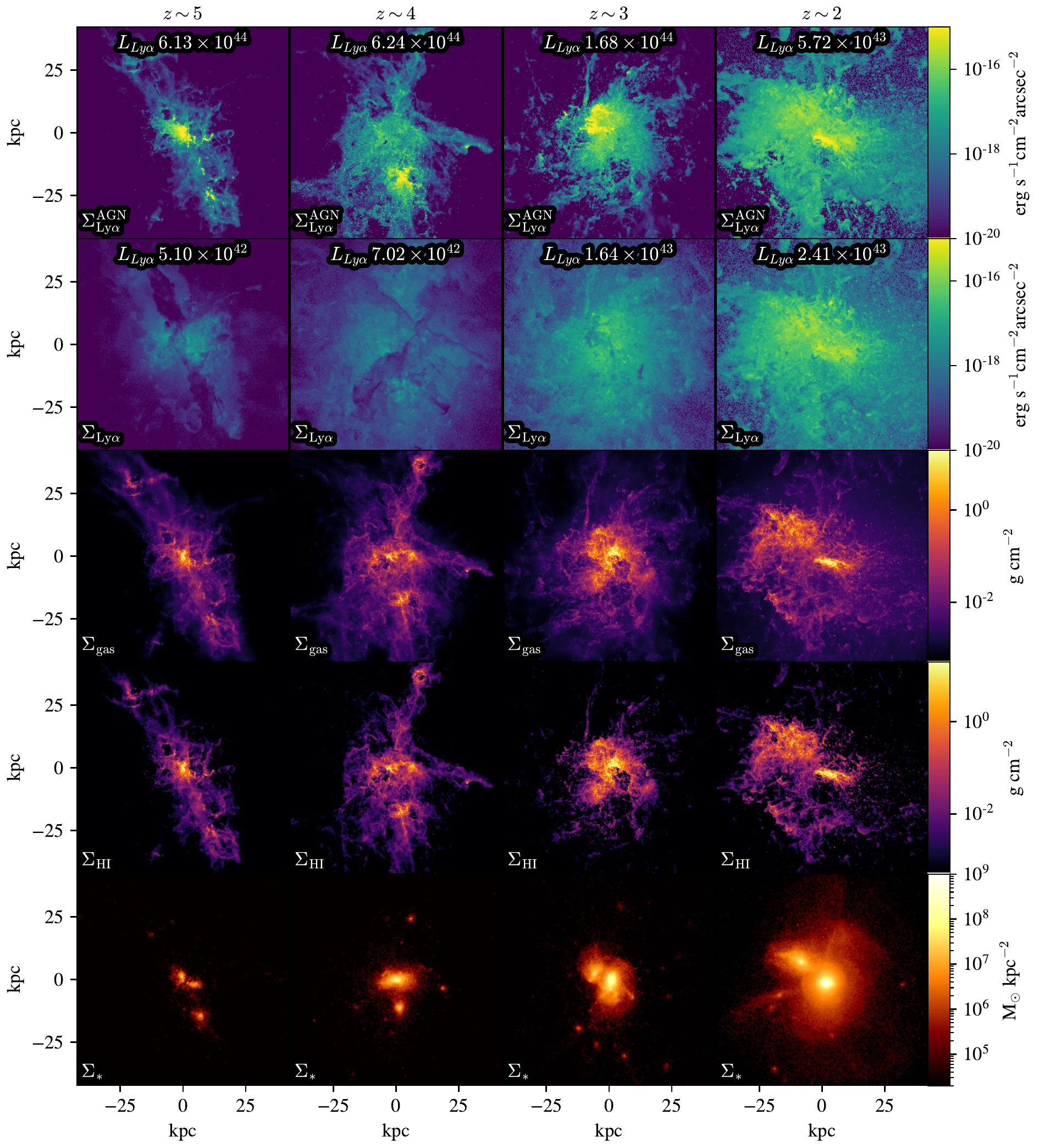}
    \caption{Ly$\alpha$ surface brightness with our AGN model (top row), Ly$\alpha$ surface brightness without our AGN model (second row), gas surface density (third row), and stellar surface density (bottom row) at four redshifts for galaxy A4. All images are 75$\times$75 physical kpc across, corresponding to 11.7, 10.5, 9.5, and 8.7 arcseconds across, respectively. Each row shares the same color scale. Though our Ly$\alpha$ RT code produces spectra, we defer the analysis of these to another paper. \aarono{Mark AGN locations in top row (point or X).}}
  \label{fig:grid_plot}
\end{figure*}

Figure~\ref{fig:grid_plot} shows $75$\,kpc postage stamps of our fiducial model halo at
integer redshifts between $z = 2$--$5$.  We show the total Ly$\alpha$
luminosity, Ly$\alpha$ surface brightness, gas surface brightness, and
stellar surface densities. We will return to
  Figure~\ref{fig:grid_plot} repeatedly throughout this paper, though
  note that large Ly$\alpha$ luminosities and extended morphologies
  are evident at a range of times in the galaxy's history.

\subsection{The formation of Ly\texorpdfstring{$\alpha$}{a} blobs}
\label{sec:formation_of_labs}

The first question we pursue is whether our simulations can actually
form a Ly$\alpha$ blob.  Recalling \S~\ref{section:blob_definition},
there is no formal definition of what constitutes a LAB.  We therefore
explore two reasonable criteria for objects classified as Ly$\alpha$
blobs in comparison to our simulations.

\begin{figure}
  \centering
  \includegraphics[width=\columnwidth]{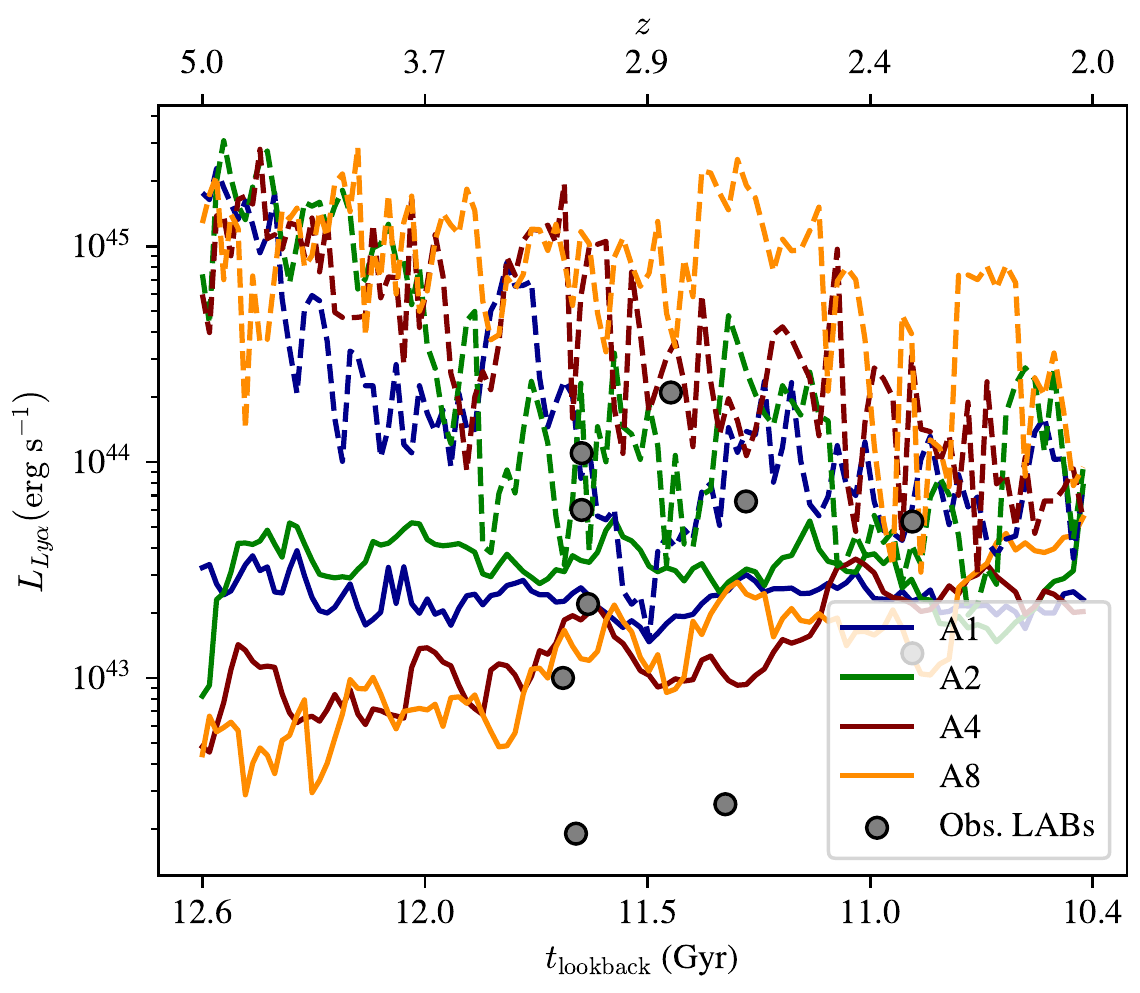}
  \caption{
      Ly$\alpha$ luminosity (median over all sightlines) for each galaxy in our sample without (solid lines) and with (dashed lines) AGN, alongside observational data from Table~\ref{table:labs} overplotted as gray points.
    Broadly, our LABs fall within the range of observed objects between $z = 5$ and $z = 2$.
  }
  \label{fig:luminosity_redshift}
\end{figure}

We first consider a total Ly$\alpha$ luminosity-based definition.  In Figure~\ref{fig:luminosity_redshift}, we plot the Ly$\alpha$ luminosity for our model galaxies as a function of time from \mbox{$z \approx 2$--$6$}.  For comparison, we also show the Ly$\alpha$ luminosities for a number of observed LABs mentioned in Table \ref{table:labs}.  The Ly$\alpha$ luminosity of our model galaxies varies substantially, but broadly overlap  with the observed range of luminosities over the considered redshift range.

\begin{figure}
    \centering
    \includegraphics[width=\columnwidth]{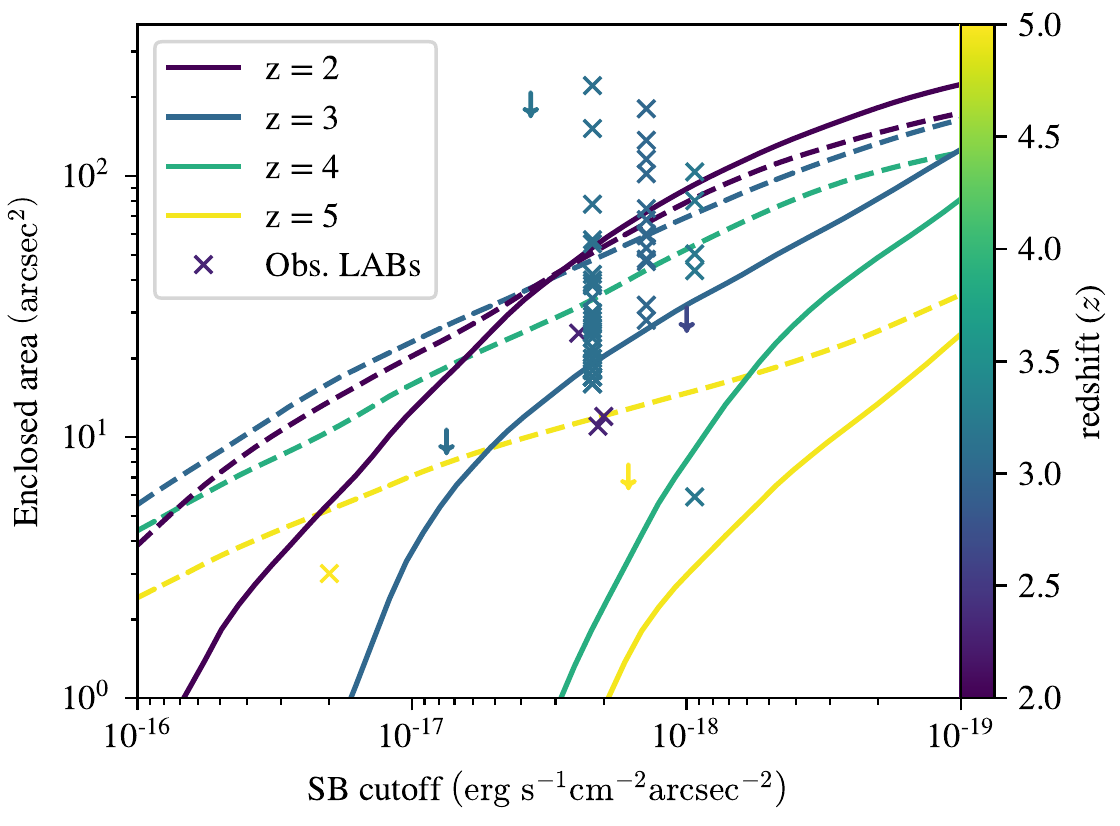}
    \caption{
        Comparison of blob sizes in our models to literature sizes.
        We define the size as the area enclosed with in a surface brightness contour, and plot models that both include AGN (dashed lines) as well as those that don't (solid lines).
        Though this plot is made in square arcseconds as opposed to physical kpc, angular size varies by less than a factor of 2 over the redshift range we study ($2<z<5$), which is not substantial on these axes.
    }
    \label{fig:area_plot}
\end{figure}

At the same time, LABs are known not only for their prodigious
Ly$\alpha$ luminosity, but also their extended morphologies. Some
studies therefore employ surface brightness profiles to characterize
the spatial extent of objects \citep[e.g.][]{Wisotzki2018}. However,
as we will demonstrate, the blob morphologies are sufficiently
disordered and asymmetric that it's not entirely obvious how to define
a radial profile. In order to characterize blobs by their surface
brightness, in Figure~\ref{fig:area_plot}, we plot the area enclosed
by a number of isophotes as a function of the isophotal luminosity. We
also attempt to compare these to known LABs (from Table
\ref{table:labs}). Since we want to plot an area and most publications
only mention a radius or semi-major axis of a blob, we assume such
observed blobs are circular to deduce an area, and therefore denote these as
upper limits since the true beam filling factor is likely
lower than unity. This comparison to LABs is preferable to a surface brightness
profile because while both collapse the azimuthal dimension to assist
in easy comparison between objects, a surface brightness profile
typically assumes azimuthal symmetry, which is typically not the case
for LABs.

\begin{figure*}
    \centering
    \includegraphics[width=2.1\columnwidth]{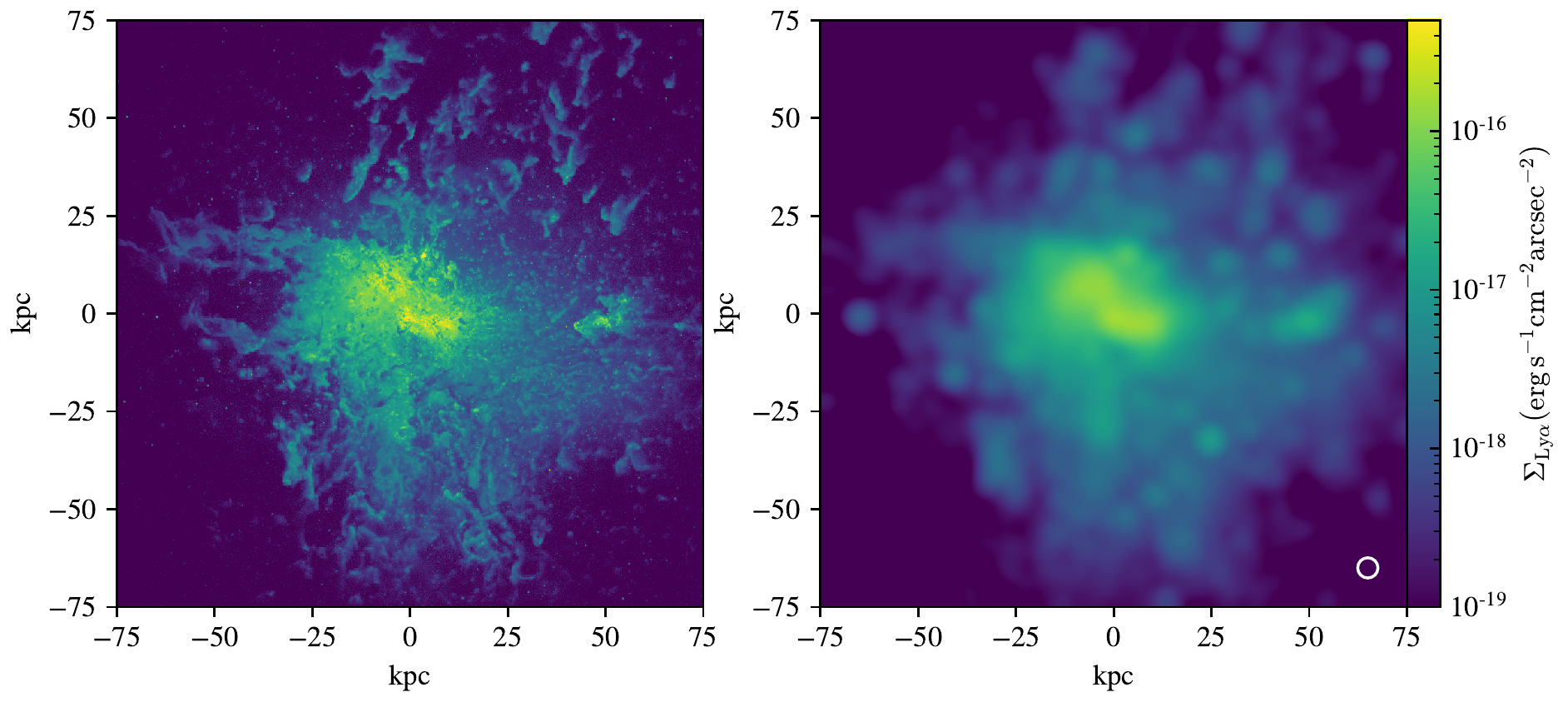}
    \caption{
        In the left panel we show one of our surface brightness images at a very high resolution, and in the right panel we convolve the surface brightness down to the $0.6''$ resolution at $z=2$ ($2.5$ kpc) of MUSE with Gaussian noise at $\sigma = 10^{-18}$\,erg/s to produce an image that more closely resembles current observations of LABs.
        \aaron{Add more info to the figure, e.g. redshift, ``(No AGN)'', scale for physical size}
    }
    \label{fig:MUSE}
\end{figure*}

As is evident from
Figures~\ref{fig:luminosity_redshift} and ~\ref{fig:area_plot}, our
model galaxies display reasonable Ly$\alpha$ luminosities and enclosed
areas as a function of limiting surface brightness when compared to
observations. In Figure~\ref{fig:MUSE}, we take a
$z \sim 2$ snapshot of our fiducial model and convolve the model
Ly$\alpha$ surface brightness with the point source function (PSF) of the Multi Unit Spectroscopic Explorer (MUSE) at the Very Large Telescope (VLT).  We note that the observed morphology when convolved with observed PSFs resembles the observations.

We now spend the bulk of the remainder of this paper unpacking
Figures~\ref{fig:luminosity_redshift} and \ref{fig:area_plot},
exploring {\it why} these galaxies emit copious Ly$\alpha$ emission.

\section{Origins of Observed Ly\texorpdfstring{$\alpha$}{a} Photons in Giant Blobs}
\label{sec:origins}
In this section, we conduct a series of numerical experiments in order
to characterize the driving sources of Ly$\alpha$ radiation from our
model blobs. We investigate the relative contributions of emission
from gas cooling and recombinations
(Figures~\ref{fig:recombination_collision}~and~\ref{fig:agn_recombination_collision}), the impact of the
ionizing UV background (Figure~\ref{fig:uvb_morphology}), and the
presence of AGN (Figure~\ref{fig:agn_comparison}). We find
that our model LABs can be powered by a combination of recombination
in star-forming galaxies, as well as cooling from accretion, which we
define as emission from collisionally excited neutral hydrogen. When
we include a model for the influence of AGN, this also contributes
significantly to the LAB luminosities. As we will show, the relative
contribution to the total Ly$\alpha$ power from each emission source
varies strongly over cosmic time, reflecting the diverse physical conditions that occur during massive galaxy evolution.

\subsection{Basic Physical Concepts}
\label{sec:physicalconcepts}
We first discuss the basic physics driving Ly$\alpha$ emission from
cooling gas and emission from ionized hydrogen recombining in a
parcel of gas before applying these insights to our model galaxies.
The physics controlling emission from these two sources is coupled;
consequently, we discuss emission from cooling gas and recombinations
simultaneously.

Cooling emission is primarily driven by gas accretion onto the central galaxy, and produces Ly$\alpha$ emission by collisionally exciting neutral hydrogen with free electrons. The rate of cooling emission is therefore proportional to the product of neutral hydrogen and free electron densities:
\begin{equation}
\label{eq:j_col}
    L_{\text{Ly}\alpha}^\text{col} = h\nu_{\alpha} \int C_{1s2p}(T)\,n_e\,n_\text{\HI}\,\text{d}V \, ,
\end{equation}
where $C_{1s2p}(T)$ is the temperature-dependent collisional rate coefficient \citep{Scholz1991}, and has units of $\text{cm}^3\,\text{s}^{-1}$.
$h\nu_{\alpha}$ denotes the energy of a Ly$\alpha$ photon, $n_e$ the number density of electrons,  and $n_\text{\HI}$ the number density of neutral hydrogen. Since we mostly deal with environments that have high ionization fractions, the free electron number density is approximately equal to the number density of ionized hydrogen, and thus the collisional excitation is maximized where approximately half the hydrogen is ionized.

While emission from collisional excitation is driven by the presence of \HI\ and free electrons, star formation produces Ly$\alpha$ by case-B recombination in heavily ionized regions. These recombinations emit Ly$\alpha$ at a rate which is proportional to $n_e n_\text{\HII}$, given by:
\begin{equation}
\label{eq:j_rec}
    L_{\text{Ly}\alpha}^\text{rec} = h\nu_{\alpha} \int P_B(T)\,\alpha_B(T)\,n_e\,n_\text{\HII}\,\text{d}V \, ,
\end{equation}
where $P_B(T)$ is the Ly$\alpha$ conversion probability per recombination event  and $\alpha_B(T)$ is the case-B recombination coefficient \citep{Cantalupo2005, Dijkstra2014, Hui1997}.

\begin{figure}
  \centering
  \includegraphics[width=\columnwidth]{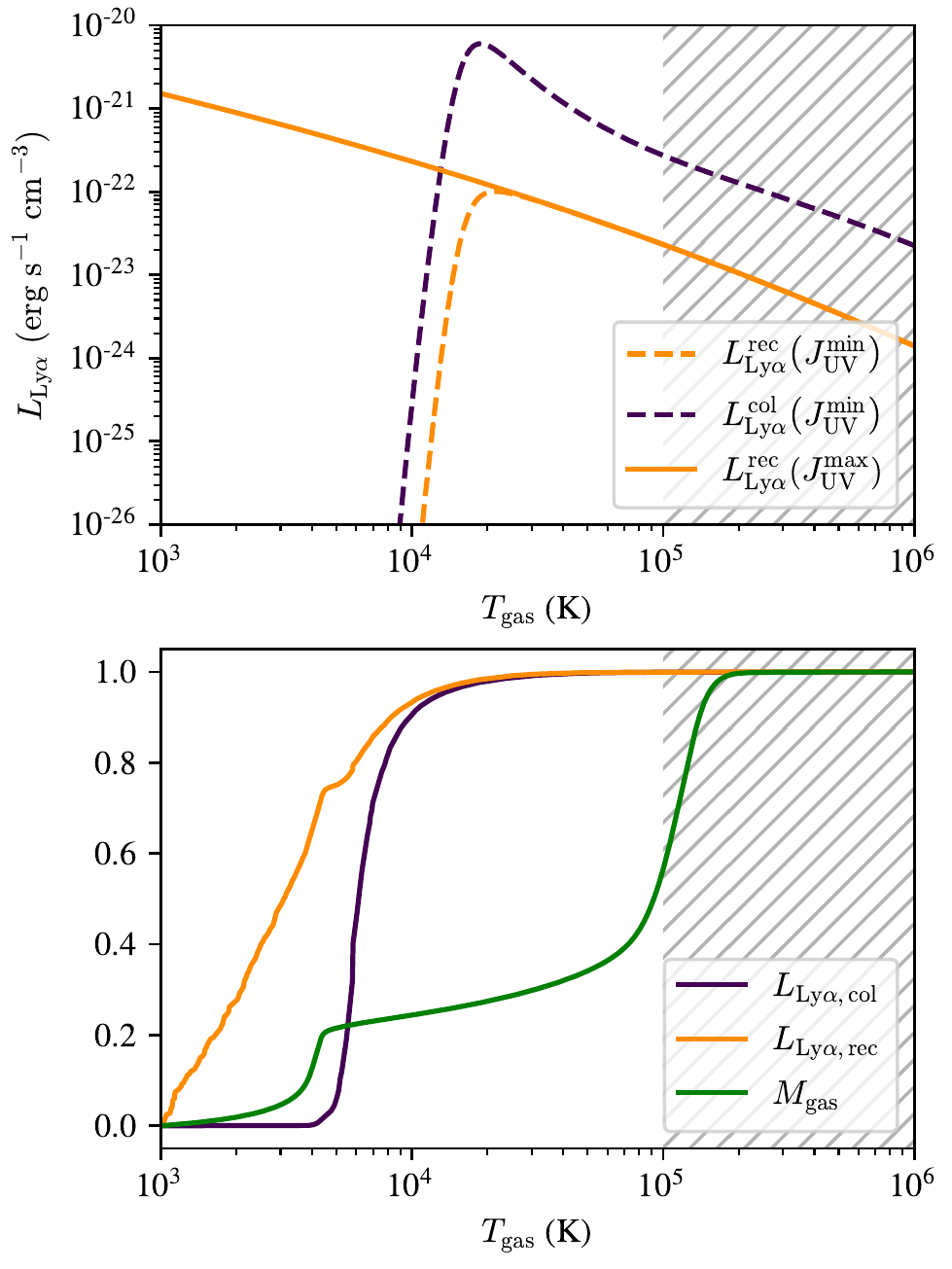}
  \caption{
    In the top panel, we show the Ly$\alpha$ luminosity for a parcel of gas at an average density and metallicity in our simulations as a function of temperature.
    The dashed lines represent the luminosity of the fiducial gas at a minimal UV field, and the solid line at the maximal UV field.
    In both panels, the purple curves represent Ly$\alpha$ emission due to collisionally excited neutral hydrogen, and the orange curves represent emission due to hydrogen recombination.
    The gray hatched region for $T > 10^5$ K indicates a temperature range for which we have extrapolated the Ly$\alpha$ conversion probability per recombination event as these probabilities are not computed at $T>10^5$ K in the \citet{Pengelly1964} tables that we utilize.
    In the bottom panel, we show the cumulative distribution of escaping Ly$\alpha$ by source, with respect to the temperature of the gas it is emitted from.
    In our simulations, we find that the Ly$\alpha$ emission from recombination traces a cooler population of gas than the emission due to collisional excitations.
    But we also observe that there is a substantial quantity of gas which does not emit strongly at all (because it is too hot or diffuse) and which does not participate in Ly$\alpha$ scattering (because it is fully ionized by its elevated temperature).
    \aarono{This caption is too long, find a way to be more concise (refer to footnote, etc).}
  }
  \label{fig:luminosity_vs_temperature}
\end{figure}

To demonstrate the relationship between the sources of Ly$\alpha$
emission and gas physical conditions, in Figure
\ref{fig:luminosity_vs_temperature}, we set up a controlled idealized
experiment in which we bathe a $1\,\text{cm}^{3}$ cube of gas at the median density in our simulations in a radiation
field with intensity $J_\text{UV}$, and plot two limiting cases for the
luminosity of this specific volume of gas as a function of
temperature: with low $J_\text{UV}^\text{min} = 0\ \text{erg\ s}^{-1}$
and high $J_\text{UV}^\text{max} = 10^{-5}\,\text{erg\ s}^{-1}$. The
high-UV value was chosen to fully ionize the gas\footnote{Note that
in Figure \ref{fig:luminosity_vs_temperature} we plot the Ly$\alpha$
luminosity across the full temperature range seen in our
simulations, but the analytical approximation we use for for
$P_B(T)$ only extends out to $10^5$ K because that is the limit of
the tables in \citet{Pengelly1964}. Therefore we have shaded this
region of the plot to indicate that this region where
$P_B\left(T\right)$ is being extrapolated from the analytic
formulation. It is likely these tables do not extend to very high temperatures because hydrogen will be mostly collisionally ionized (depending on the density)}.

We first consider the $J_\text{UV}^\text{min}$ case in
Figure~\ref{fig:luminosity_vs_temperature} (purple). Here, emission is
maximized near $T=10^4$ K, because the impact of the gas temperature
on collisionally-driven Ly$\alpha$ emission is twofold. In the very
low temperature regime, the ionization rate is sufficiently low that
there are no free electrons to collisionally excite the gas. As the
ionized fraction increases with temperature, there are more free
electrons but less neutral hydrogen to be collisionally
excited. However the second effect of temperature is to increase the
rate of collisions, which produces a strong mitigating effect against
the dropping abundance of neutral hydrogen; even as the gas approaches
being fully ionized at high temperatures the rate of collisions
mitigates the drop in luminosity.

Turning now to the $J_\text{UV}^\text{max}$ case in our idealized
numerical experiment (top panel of
Figure~\ref{fig:luminosity_vs_temperature}, orange lines) the
Ly$\alpha$ emission from recombination declines slowly with
temperature. The $J_\text{UV}^\text{max}$ also has a different
temperature dependence. The gas is maximally ionized at all
temperatures, but we see a decline in emissivity with temperature
because the cross-section of the electrons and hydrogen nuclei drop
as their thermal velocities increase, making recombination less likely. Note that in this
experiment, where the the UV field should completely ionize the
hydrogen, it is prevented from doing so to preserve numerical
stability: the neutral fraction is restricted from dropping below
$10^{-10}$.  As a result, in this extreme scenario, the collisional
emissivity is driven by this neutral fraction floor, and is therefore
unphysical.  Accordingly, we do not plot the collisional emissivity in
the $J_\text{UV}^\text{max}$ case in
Figure~\ref{fig:luminosity_vs_temperature}.
The luminosity due to collisional excitations whcih is produced by this numerical artefact is many orders of magnitude below the luminosity due to recombinations.
Removing it would not alter the results presented in this work.

The trends in our controlled experiment
(Figure~\ref{fig:luminosity_vs_temperature}) provide us with the
physical insight we need to understand which gas in our simulation is
emitting Ly$\alpha$, and which gas is not. In the bottom panel of
Figure~\ref{fig:luminosity_vs_temperature}, we show the cumulative
distribution of recombination and collisionally-excited emission in a
single snapshot. From this we can see that the bulk of the Ly$\alpha$
photons are produced by ``cold'' photoionized gas gas ($T < 5\times
10^3$ K) and the ``warm'' collisionally-excited gas $(6\times10^3\ \text{K} \leq T \leq 10^4\ \text{K})$.
The emission sources (recombination and collisional de-excitation) are segregated by temperature on account of the astrophysical mechanisms responsible for the gas temperature.
Cooler, recombining gas gas typically lies at high densities with efficient cooling and must be ionized primarily by a nearby UV source, i.e. newly formed stars.

\subsection{Ly\texorpdfstring{$\alpha$}{a} Emission from Cosmological Simulations of Massive Galaxy Evolution}
\label{sec:emission_in_simulations}
Now that we have built insight into the physics of Ly$\alpha$ emission
from collisional excitation and recombination in an idealized
experiment, we turn to our galaxy evolution simulations to understand
the dominant sources of Ly$\alpha$ luminosity in our model LABs.

\begin{figure}
  \centering
  \includegraphics[width=\columnwidth]{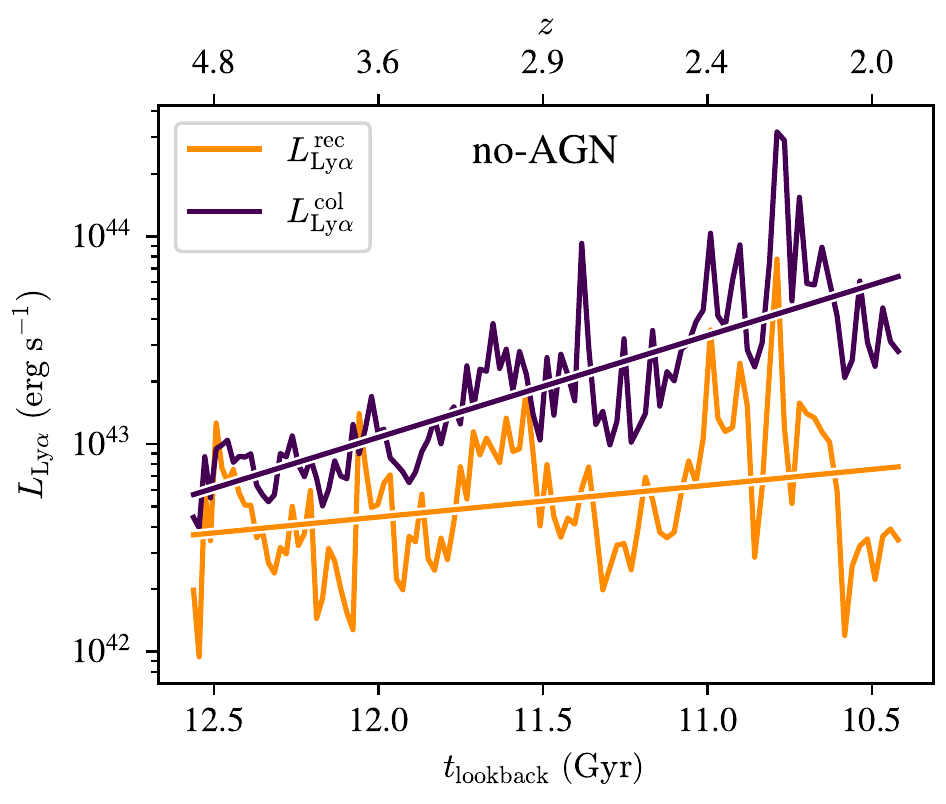}
  \caption{Ly$\alpha$ luminosity from our fiducial LAB broken down by
    source of emission as a function of redshift. \aarono{It would be nice to combine this figure and figure 8, as a two panel plot, as they share the same time/redshift axes and would make it easier to directly compare with and without AGN. (Even though this is orange, I think it would be best.)} \aarono{Give the best fit line (or power-law), e.g. as $\log L_{\text{Ly}\alpha} \approx L_{\text{Ly}\alpha,0} (1+z)^\beta$.}}
  \label{fig:recombination_collision}
\end{figure}

\begin{figure}
  \centering
  \includegraphics[width=\columnwidth]{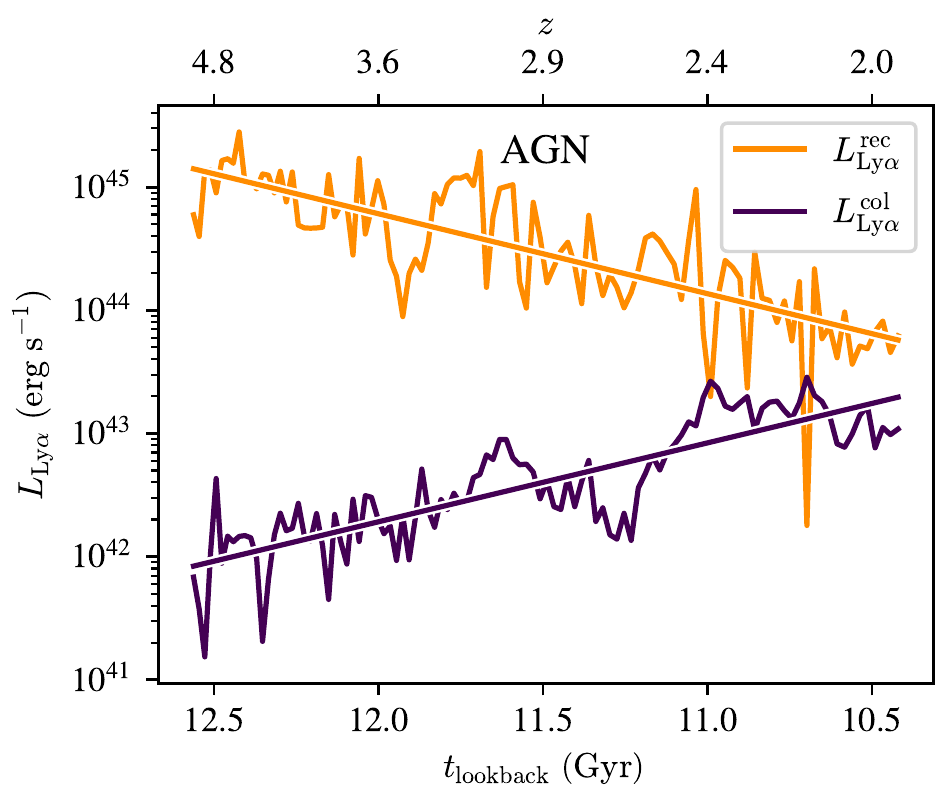}
  \caption{Galaxy A4 emission broken down by source of emission over redshift; as opposed to Figure~\ref{fig:recombination_collision} this plot includes the effect of AGN. \aarono{Combine with figure 7? Okay I see, it is fine as is.}\ben{Originally they were combined, but we split the figure because AGN has its own section so we wanted to have all the AGN-specific material separated. That goal has become somewhat muddied since then.}}
  \label{fig:agn_recombination_collision}
\end{figure}

In Figure \ref{fig:recombination_collision} we plot
independently the recombination and collisional excitation components
of our fiducial LAB's luminosity. The contributions from recombination
and collisions vary dramatically over the course of the model halo's evolution
evolution, though by and large emission from collisionally excited
hydrogen dominates, and grows over redshift as this galaxy grows.
Integrating over our redshift of interest ($2 \leq z \leq 5$) we find $\frac{\int L_{\text{Ly}\alpha}^\text{col} \text{d}t}{\int
  L_{\text{Ly}\alpha}^\text{col} + L_{\text{Ly}\alpha}^\text{rec} \text{d}t} =
0.80$. In \S~\ref{sec:agn} we will discuss the
impact of including an AGN in these models; when we include AGN
recombination from emissions dominates and the above ratio becomes $0.03$.

\begin{figure*}
  \centering
  \includegraphics[width=2.1\columnwidth]{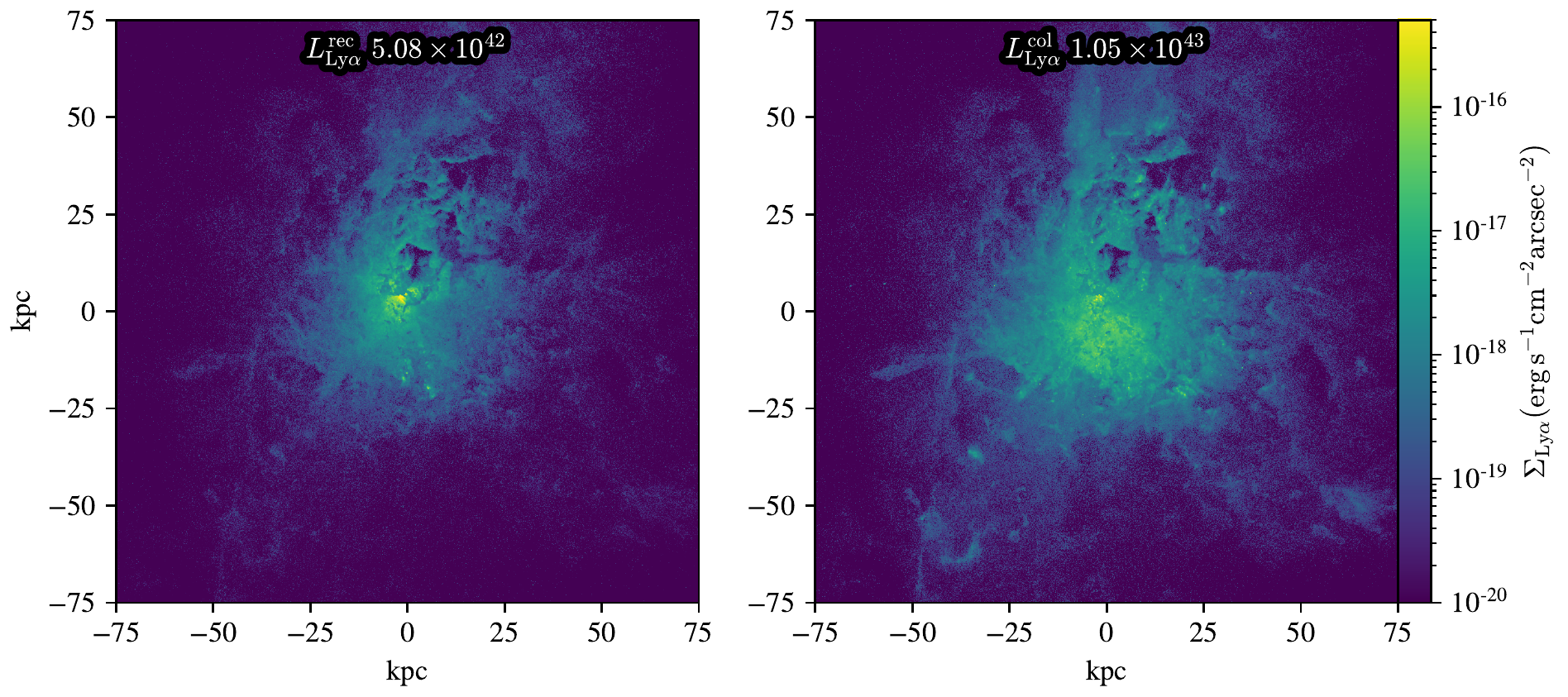}
  \caption{The $z=3.0$ snapshot from galaxy A4 with only emission from recombinations, in the left panel, and only emission from collisional excitation on the right panel. Note that the emission from recombination and emission from collisional excitation is distributed differently across the halo.}
  \label{fig:121_rec_col}
\end{figure*}

It is tempting to ask whether the dominant power source (i.e. recombinations vs collisions) are correlated with an obvious physical property of the galaxy?
Across our three galaxies, we do not discover any strong trends (plots of this non-result can be found in Appendix~\ref{app:correlations}).
The reason for this is complex: as we demonstrated in Figure~\ref{fig:luminosity_vs_temperature}, the relative contribution
of recombinations and collisions is a complex function of both the gas
temperature and incident radiation field on a given parcel of
gas. Galaxies have a large distribution of temperatures and ionization
states that vary over the course of their lifetimes, and this
distribution does not vary smoothly with a single physical
property. The radiation fields are dependent on the small scale
clumping and opacity variations across the galaxy, which result in the
dominant power source (recombinations vs collisions) varying
non-monotonically across the galaxy. We can see this explicitly in
Figure~\ref{fig:121_rec_col}, where we show the morphology of galaxy
A4 at redshift $z=3$ while isolating the recombination driven and
collisionally driven luminosity, respectively. The former naturally
peaks at the center of the galaxy, where AGN and star formation-driven
ionization peaks.  However, emission from both physical processes is
significant across the bulk of the main disk.

\section{The Impact of AGN on Ly\texorpdfstring{$\alpha$}{a} Emission from Massive Halos}
\label{sec:agn}
We now turn our attention to the influence of an AGN on our modeled Ly$\alpha$ emission.
We note that we do not explicitly include AGN feedback; instead, from the perspective of the hydrodynamic simulations, black holes are included as passive sink particles that only accrete (\S~\ref{sec:methods}).
This said, we are able to assess their impact on the emission properties of the simulations in postprocessing.
Here, we treat AGN as an ionizing source when we compute the ionization state of the gas with {\sc lycrt}.
In this model, the AGN SED is modeled by employing the \citet*{Hopkins2007} templates for unreddened quasars, with the luminosity being tied only to the mass of the black hole particle (as opposed to the accretion rate in the simulation) by assuming the black hole is always accreting at the Eddington rate, with an efficiency of $\eta = 0.1$
\footnote{
    It should be noted that this is an overly simplistic model of AGN, neglecting departures from isotropic radiation, accretion rate variability, and the impact of radiative and mechanical feedback on the thermodynamic state of the gas.
}.
In what follows, we investigate the impact of AGN on the total Ly$\alpha$ luminosity, as well as the overall spatial extent of the blob.

\subsection{Impact of AGN on Luminosity and Escape Fraction}
\begin{figure}
  \centering
  \includegraphics[width=\columnwidth]{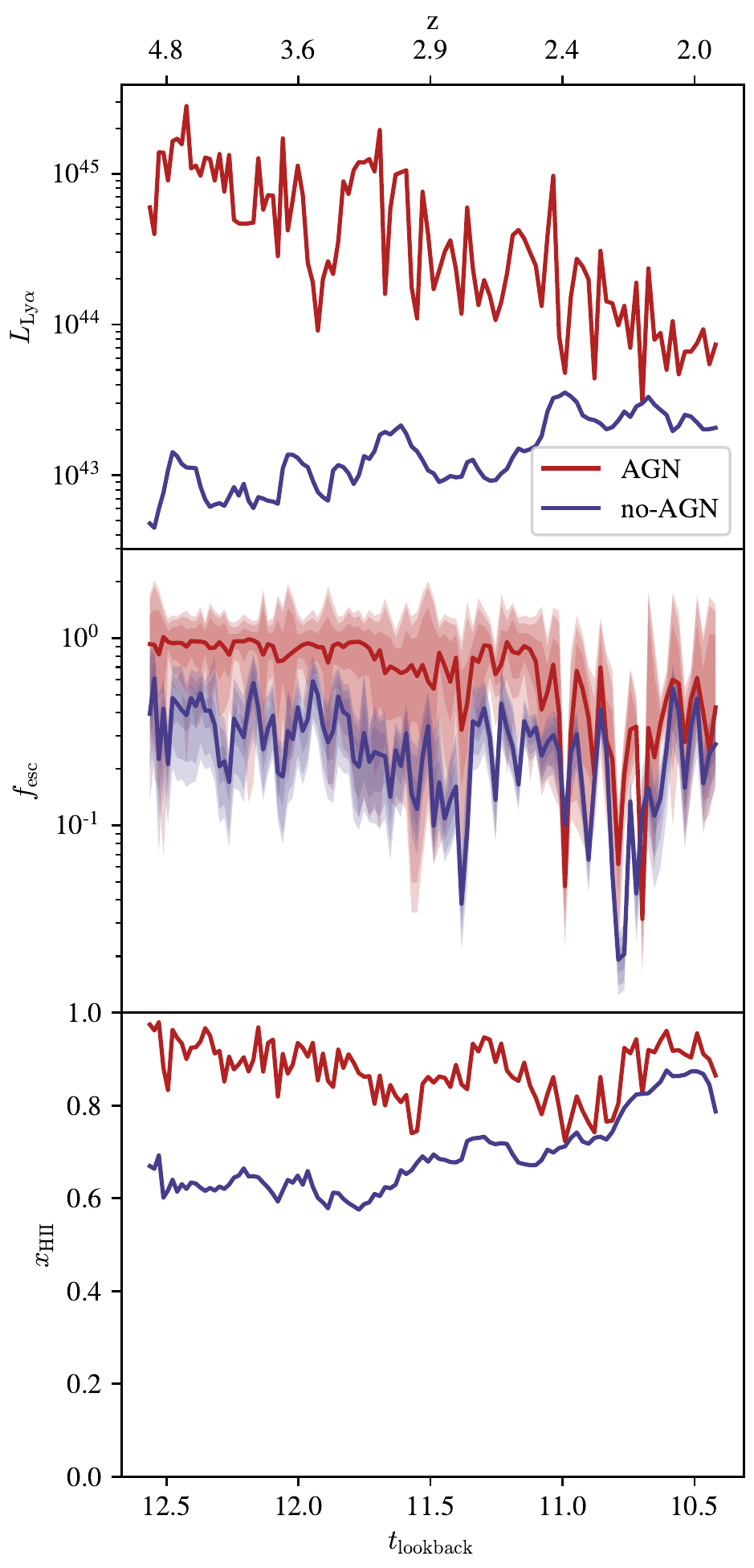}
  \caption{
  We compare the Ly$\alpha$ luminosity, Ly$\alpha$ escape fraction, and ionization state of the galaxy and halo with and without an AGN model.
  In the escape fraction plot the shaded regions represent $1\sigma$, $2\sigma$, and $3\sigma$.
  The luminosity of a Ly$\alpha$ blob can be substantially enhanced by the AGN model.
  The simulation domain is always heavily ionized, but the presence of AGN also provides stochastic enhancements, though it does not well correlate with luminosity or escape fraction.
  }
  \label{fig:agn_comparison}
\end{figure}

In Figure~\ref{fig:agn_comparison}, we plot a comparison of the time evolution of the Ly$\alpha$ luminosity, Ly$\alpha$ escape fraction, and ionized gas fraction for models with and without an AGN for our fiducial galaxy.  As is evident, there are significant differences in a model that includes AGN compared to one that does not\footnote{The snapshots this work is based on are not sufficiently high-resolution to capture some small-scale clumping in the multiphase ISM, and therefore we may be overestimating the ionization of the gas in general but specifically in the presence of an AGN due to a lack of small self-shielded clumps.}.

\begin{figure}
    \centering
    \includegraphics[width=\columnwidth]{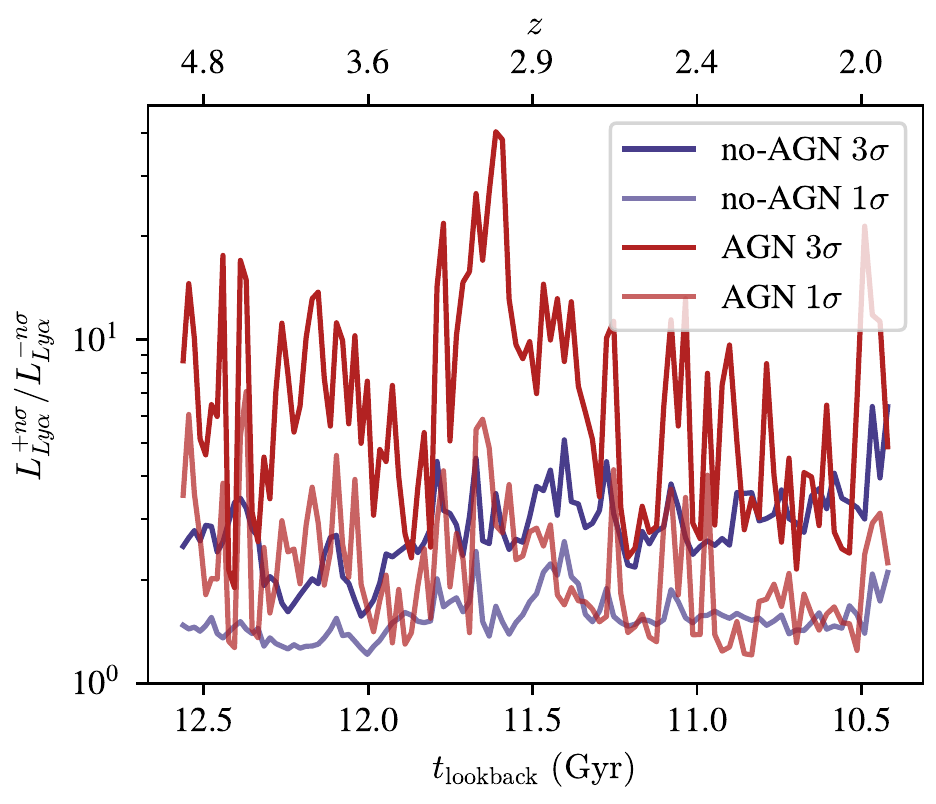}
    \caption{
        Maximum fractional difference in luminosity (or escape fraction) between lines of sight over redshift for one of our halos that forms a LAB, with and without AGN.
        The value on the y-axis is notionally by what factor the luminosity of the object varies by with respect to different lines of sight.
    }
    \label{fig:los_variation}
\end{figure}

Since Ly$\alpha$ escape is sightline-dependent we show in the second panel of Figure~\ref{fig:agn_comparison} the variation of our fiducial LAB's escape fraction (and threfore luminosity) over sightlines, with ionization due to AGN and without.
The observed luminosity can vary substantially due to the viewing angle of the galaxy when AGN are present.
To demonstrate this explicitly, in Figure~\ref{fig:los_variation}, we plot the 3-$\sigma$ relative variation between sightlines to show how different a single physical object may appear to an observer who can only view the object from one line of sight.
We use 3$\sigma$ as a way to quantify the range of possible observed values, but since we only use 3072 lines of sight to compute the related percentiles, the 3$\sigma$ values are sensitive to only $\sim 10$ lines of sight.
Therefore we include the 1$\sigma$ quantities as well because though they do not have quite the same meaning, they closely resemble the 3$\sigma$ quantities which indicates the darker 3$\sigma$ curves are not extremely sensitive to outliers.
This huge variation when AGN are present is caused by the distinct non-uniformity of CGM opacity; the AGN is able to punch large ionization holes in the enclosing CGM through which Ly$\alpha$ readily escapes.
It is important to note that the distribution of escape fractions is not normal; we sample 3072 sightlines to produce the middle panel of Figure~\ref{fig:agn_comparison} which is sufficient to explore many of the high-escape pathways out of a LAB.
Because of this sightline variation, an escape fraction (or luminosity) calculated along a single line of sight may not be particularly representative of the overall galaxy properties.
However, individual sightlines are still meaningful to probe the representative statistics of observed galaxy populations, especially because the observed distributions are a convolution of the sightline-independent distributions with sightline-dependence of observables.

When we include a model for AGN (which drives increased ionization in the gas), there are substantial spikes in the luminosity owing to increased emission from recombinations.  In short, the primary effects are to increase the total Ly$\alpha$ luminosity due to the increased ionization fraction of the gas. Recall from Section~\ref{sec:physicalconcepts} that as we increase the ionization state of the gas at a fixed temperature, e.g. $T \sim 10^4$\,K, the luminosity from recombinations increases while that from collisional excitations decreases.  Additionally, the escape fraction increases with the addition of an AGN (middle panel of Figure~\ref{fig:agn_comparison}), due to the fact that the Ly$\alpha$ scattering strength depends primarily on $n_\text{\HI}$ (equation~\ref{eq:kalpaha}). This escape fraction enhancement is so substantial that some orientations exhibit $f_\text{esc} > 1$ due to particular geometries that cause more Ly$\alpha$ photons to scatter into the line of sight than are absorbed.

Taken together, the increase in the ionization state of the gas (bottom panel of Figure~\ref{fig:agn_comparison}) increases both the emission from recombinations, as well as the escape fraction of Ly$\alpha$ photons. These combined effects allow for significant boosting ($\sim$ factors of $10$--$50$) of the Ly$\alpha$ luminosity compared to a no-AGN model.

\subsection{The impact of AGN on the spatial extent and concentration of Ly\texorpdfstring{$\alpha$}{a} in blobs}

As in the overall Ly$\alpha$ luminosity, the AGN can also affect the spatial extent of Ly$\alpha$ emission in massive halos. This takes two forms: (\textit{i}) the total area enclosed within a surface brightness contour and (\textit{ii}) the concentration of Ly$\alpha$ light in the system.  We explore these in turn.

Previously, in Figure~\ref{fig:area_plot}, we examined the size of our model LABs as a function of observation sensitivity (solid lines) in a fiducial model that did not have AGN on.  We now turn to the dashed lines in the same figure where we have included AGN.

We see an enhancement of blob size at high surface brightness cutoffs, which indicates that there are regions that have been substantially enhanced in brightness, but at the same time we see a decrease in blob size at much lower cutoffs. We interpret this effect as a complex interaction of the gas ionization state with escape pathways. As gas becomes more ionized it provides a pathway along which Ly$\alpha$ is likely to escape; it is these pathways that produce the small region(s) of very intense surface brightness. However, the presence of a low-opacity pathway out of the blob decreases the probability that a photon will be scattered out into the extended blob structure before it escapes.

\begin{figure}
  \centering
  \includegraphics[width=\columnwidth]{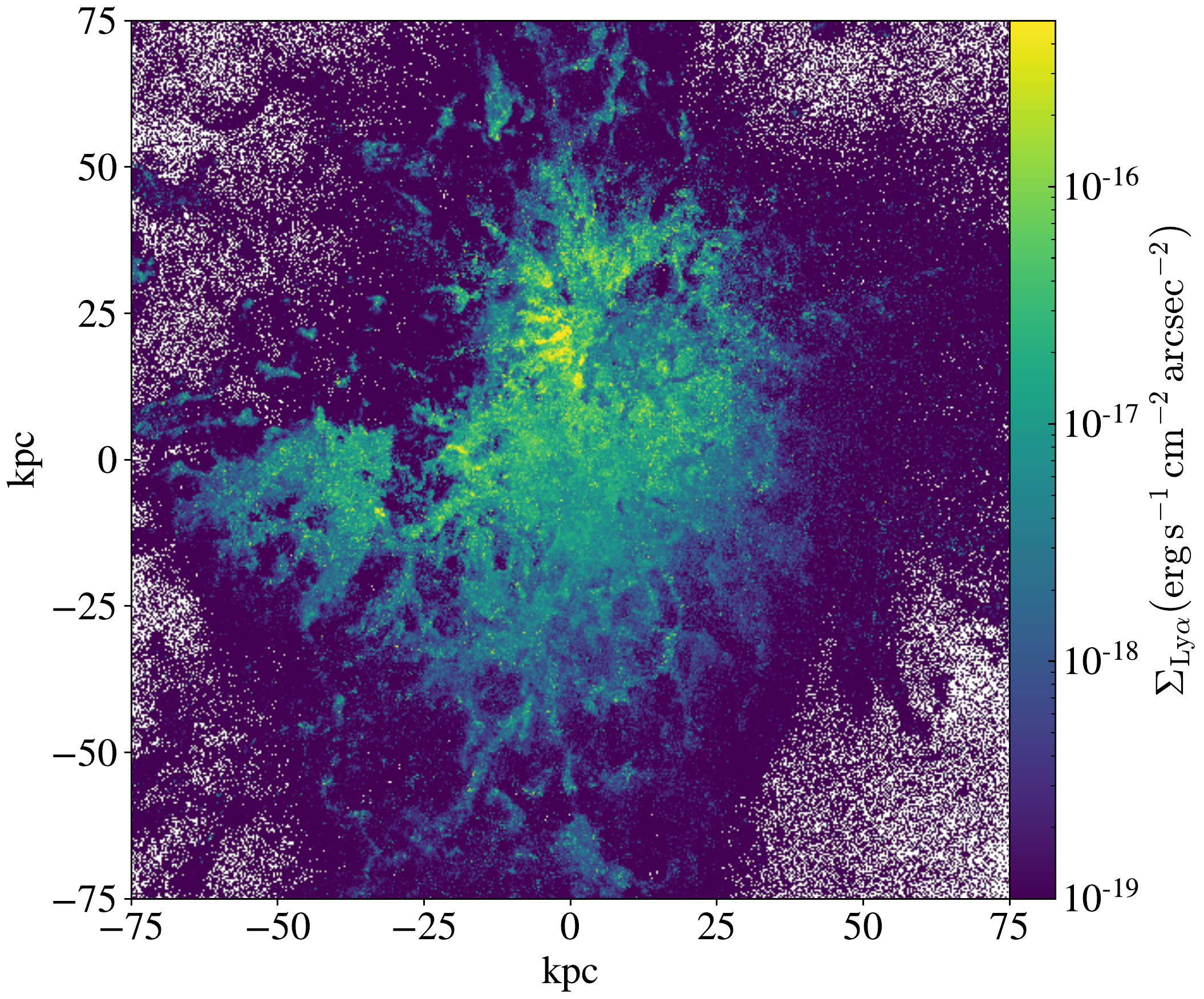}
  \caption{Example surface brightness image of a LAB where the luminosity is concentrated, which indicates the presence of an AGN, but the luminosity is not in a connected region. \aaron{Add info, redshift, etc.}}
  \label{fig:agn_on_example}
\end{figure}

\begin{figure}
  \centering
  \includegraphics[width=\columnwidth]{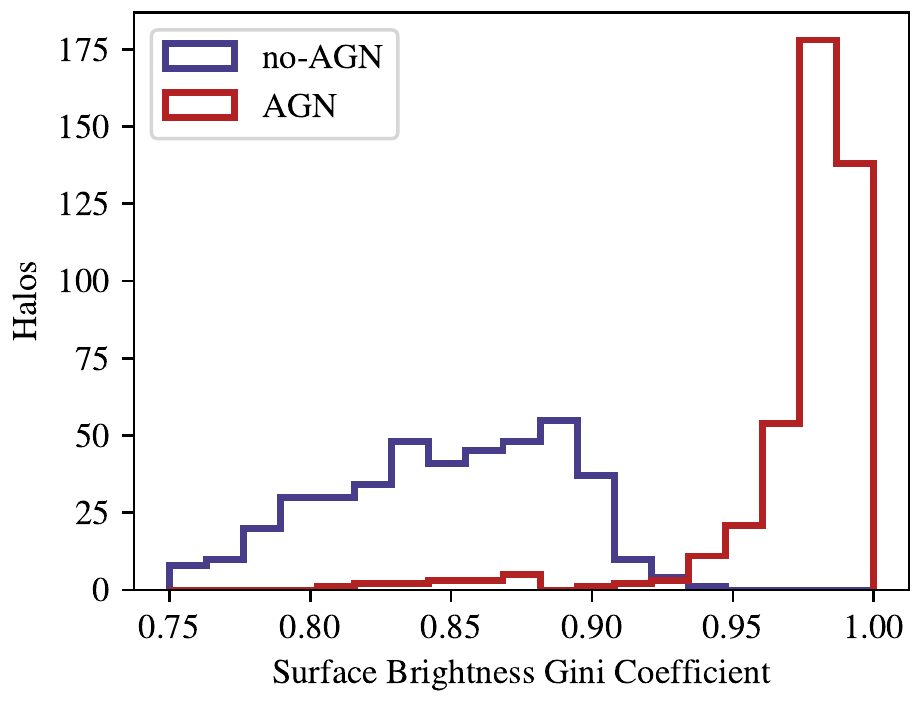}
  \caption{
      In our surface brightness images with AGN, the escaping luminosity is much more spatially concentrated.
    We quantify this trend by computing the Gini coefficient for each snapshot we have, with and without AGN, after convolving to the resolution of MUSE.
    By selecting a Gini coefficient threshold around 0.93, one could reliably distinguish between blobs that contain and do not contain AGN in our simulations. \aarono{Would this look better with a log $y$-axis?} \aarono{Is it worth adding histograms (as dashed lines) of the Gini coefficient before convolving to MUSE resolution? I.e. to guide future instruments?}}
  \label{fig:skewness}
\end{figure}

The presence of these small pathways out of the blob may be useful to detect the presence of AGN in a blob (we include a visual demonstration of this phenomenon in Figure~\ref{fig:agn_on_example}).
We quantify the concentration of light of each blob by computing the Gini coefficient of a surface brightness image defined by
\begin{equation}
    \label{eq:gini}
    G = \frac{2\sum_{i=1}^{n}i p_{i}}{n\sum_{i=1}^{n}p_{i}} - \frac{n+1}{n} \, ,
\end{equation}
where $n$ is the number of pixels and the pixel luminosity data $p_1 \ldots p_n$ are ordered by non-decreasing pixel brightness. We present histograms of the Gini coefficients for our model blobs with and without AGN in Figure~\ref{fig:skewness}. For intuition, the Gini coefficient is always in the range $G \in [0,1]$, corresponding to a range of scenarios bracketed by a uniform image ($G = 0$) to a single bright pixel ($G = 1$). The primary signature of the AGN's effect is to cause the luminosity to be concentrated in a much smaller area.
While this may seem contradictory to the increase in total luminosity and area enclosed, note that this is a {\it relative} concentration.
That is, while the diffuse emission is still significant, the central emission in the ionized bubble surrounding the AGN dominates when compared to this diffuse halo emission such that the overall concentration decreases dramatically for the AGN-on model.
This metric becomes more effective for identifying AGN with higher-resolution observations (Figure~\ref{fig:MUSE}) since the bright patches of our surface brightness images are significantly smaller than the spatial resolution of current telescopes, but current technologies should be sufficient.

We do \emph{not} propose that a Gini coefficient of specifically $0.93$ will distinguish between observations of LABs that contain an AGN and those that do not.
This metric is intended to demonstrate that it may be possible to distinguish between LABs that contain an AGN and those that do not from surface brightness features alone, and that the Gini coefficient may be an effective metric.

\section{Physical Properties of Lyman-\texorpdfstring{$\alpha$}{a} Blobs}
We have thus far investigated the origin of Ly$\alpha$ photons in massive galaxies at high-redshift, and demonstrated that they exhibit luminosities and spatial extents consistent with observed systems.  We now turn our attention to the broader physical properties of the central galaxies and their parent halos of LABs. To do this, we adopt a fiducial luminosity cutoff of $L_{\text{Ly}\alpha} \geq 10^{43}$ erg/s as a LAB.  We do not include an AGN for any of the models presented here, so our results should be considered as representing no-AGN samples.

\subsection{General Physical Properties and Thresholds}
\label{section:general_physical_properties}
\begin{figure}
    \centering
    \includegraphics[width=\columnwidth,keepaspectratio]{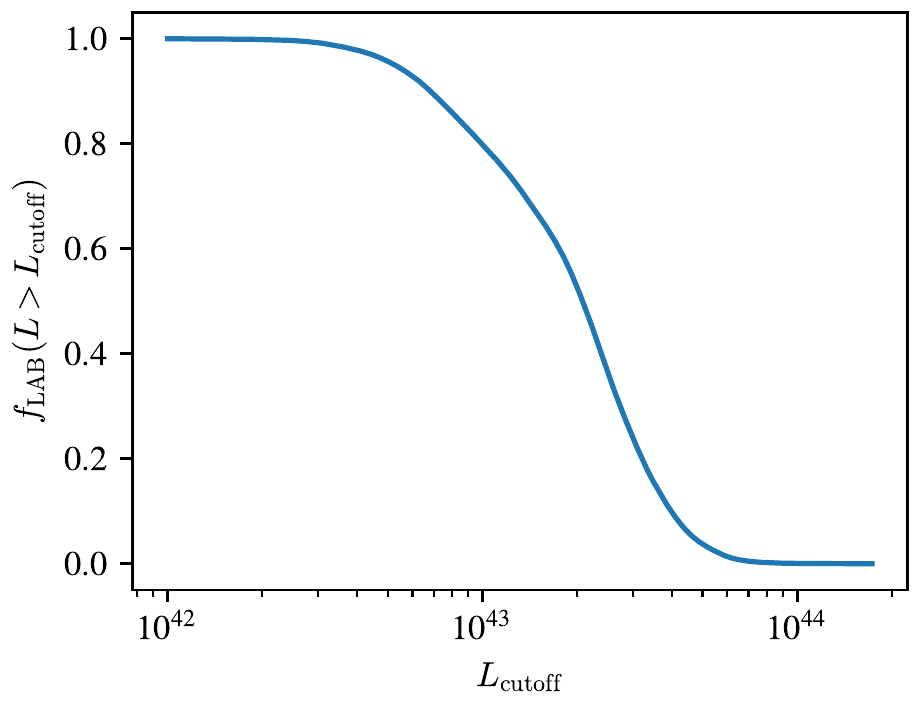}
    \caption{
        We use this distribution of the fraction of objects considered a LAB as a function of a luminosity cutoff to motivate and visualize the luminosity cutoffs presented in Figure~\ref{fig:host_properties}.
        Our 100th, 75th, 50th, and 25th percentiles are $1.72\times10^{44}$, $2.96\times10^{43}$, $2.08\times10^{43}$, and $1.18\times10^{43}$ respectively.
    }
    \label{fig:f_lab}
\end{figure}

\begin{figure*}
  \centering
    \includegraphics[width=\textwidth,height=\textheight,keepaspectratio]{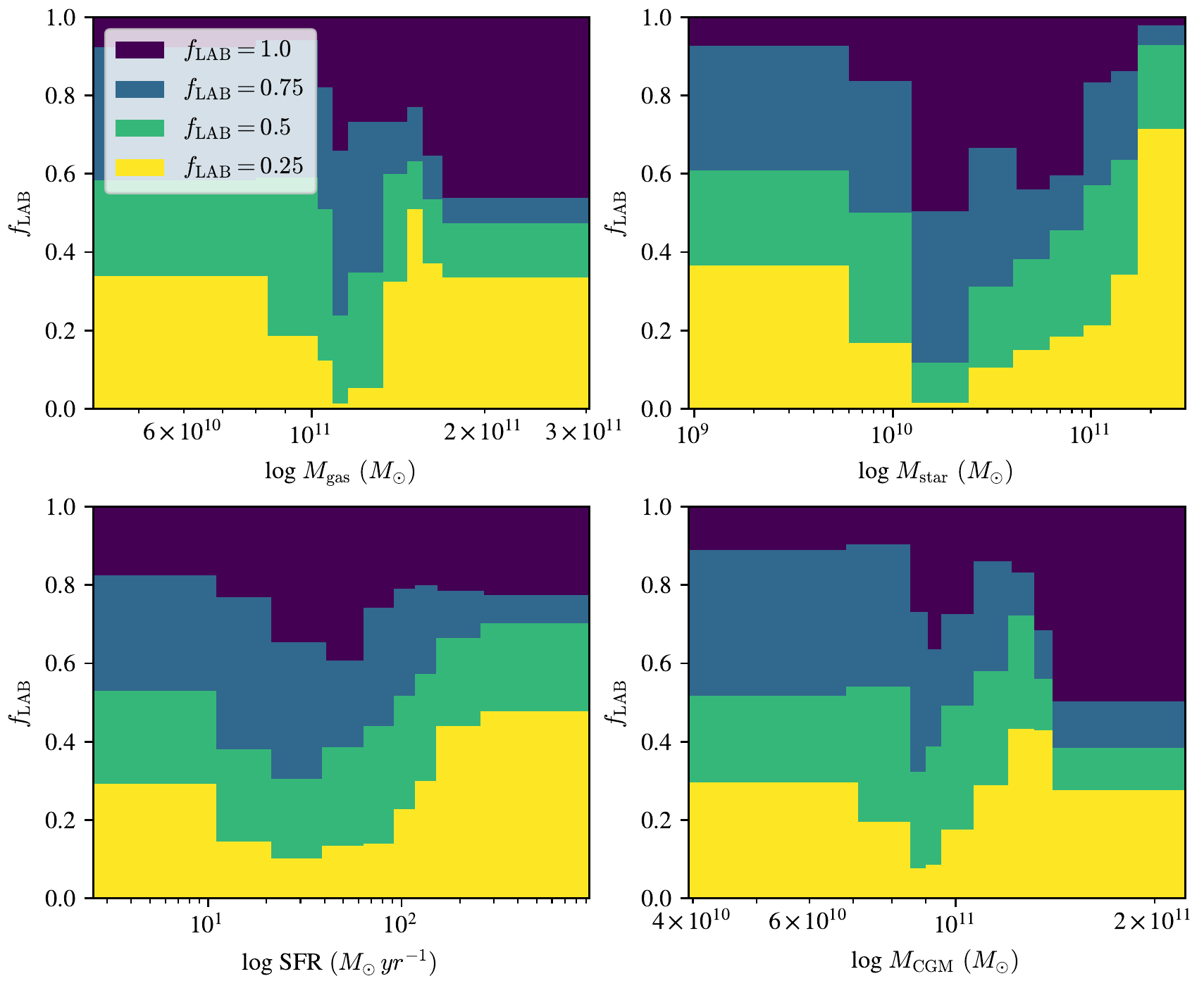}
    \caption{
        We plot the distribution of physical properties for all our snapshots at the different luminosity thresholds derived from the threshold luminosity distribution in Figure~\ref{fig:f_lab}.
        This figure compares the distribution of LAB-hosting halos at different luminosity cutoffs to demonstrate the relationship between presence of a LAB and the physical properties.
        We see a weak trend in most of the properties, but stellar mass is the strongest predictor of LAB presence.
    }
  \label{fig:host_properties}
\end{figure*}

The first question we investigate is whether a single physical property defines when a galaxy becomes a LAB.
In Figure~\ref{fig:f_lab}, we show the cumulative fraction of halos that qualify as LABs according to a luminosity threshold, i.e. $f_\text{LAB} = N(L_{\text{Ly}\alpha} > L_\text{cutoff}) / N$.
Then in Figure~\ref{fig:host_properties}, we show histograms of the gas mass, stellar mass, SFR and CGM gas mass of the simulated galaxy at all times our simulations cover, by dividing all our snapshots along luminosity divisions that equally divide the population.
The upper limits of these groups are $1.72\times10^{44}$, $2.96\times10^{43}$, $2.08\times10^{43}$, and $1.18\times10^{43}$.
We use a particular definition for the CGM gas in this work: all gas contained within our simulation box (which is approximately $4$ $R_{\text{vir}}$ on a side) which is not part of the central massive galaxy, according to an FoF group finder with a linking length set to $0.2$ times the mean interparticle separation.
Each panel of this plot contains four histograms of the halo properties selected based on luminosity thresholds that select 100\%, 75\%, 50\%, and 25\% of the total population.
On the vertical axis is the fraction of halos in the range that meet that luminosity criteria.
The general intention of this plot is to show (going from purple to yellow) how adding a more stringent luminosity cutoff alters the distribution of physical properties required to satisfy the cutoff.

In short, there are no sharp thresholds in physical properties within our modeled mass range, but we do see a general upwards trend towards greater LAB abundance with each of these physical properties, especially with $M_*$ and SFR.
But this trend is only strongly apparent at our most stringent luminosity threshold, $f_\text{LAB} = 0.25$, where the cutoff is at $L_{\text{Ly}\alpha} = 3.1\times10^{43}\,\text{erg\ s}^{-1}$, likely signifying a ``bigger things are bigger'' effect.

What is more interesting is that the gas mass and CGM gas mass are very poor predictors of whether a halo meets a luminosity cutoff.
At the outset, this may seem surprising.
Indeed in \S~\ref{sec:physicalconcepts}, we discussed the origin of Ly$\alpha$ photons in the context of gas temperature and ionization state, both properties that likely correlate with the gas mass in a halo.
This said, the physical concepts outlined in \S~\ref{sec:physicalconcepts} describe the \emph{production} of Ly$\alpha$ photons.
While important, additionally critical for the observability of Ly$\alpha$ from galaxies is the \emph{escape fraction}.
As we demonstrated in Figure~\ref{fig:agn_comparison}, there is a strong variation in the escape fraction of Ly$\alpha$ photons with viewing angle.
While the production of Ly$\alpha$ is a straightforward function of the physical properties of the gas in the simulation, the escape fractions are significantly more complex. \aarono{I realize you don't have time, but it might be interesting to check whether the intrinsic luminosity (no escape fraction effect) does behave in a more predictable way. Okay, sounds good. Can delete this comment.} \ben{I'm pretty sure I looked into this at one point and it doesn't.}

The escape fraction from a massive halo depends on the covering fraction of neutral hydrogen and dust.
Resonant scattering can either enhance the escape fraction by redirecting Ly$\alpha$ away from and around dusty media, or it can reduce the escape fraction by lengthening the path to escape, which increases the absorptive optical depth.
The worst case for escape is when the emission is deeply embedded in an envelope of multiphase, dusty, neutral gas, such as young star-forming regions.
This complexity, in essence, is what drives many of the computational challenges of these $3$D Ly$\alpha$ radiative transport calculations.

\subsection{Spatial extent of model Lyman-\texorpdfstring{$\alpha$}{a} Blobs}

We have already discussed the spatial extents of our model LABs in general, as well as in the context of observations in Figure~\ref{fig:area_plot}. Here, we simply aggregate these findings for completeness when discussing the physical properties of blobs.

\begin{figure}
  \centering
  \includegraphics[width=\columnwidth]{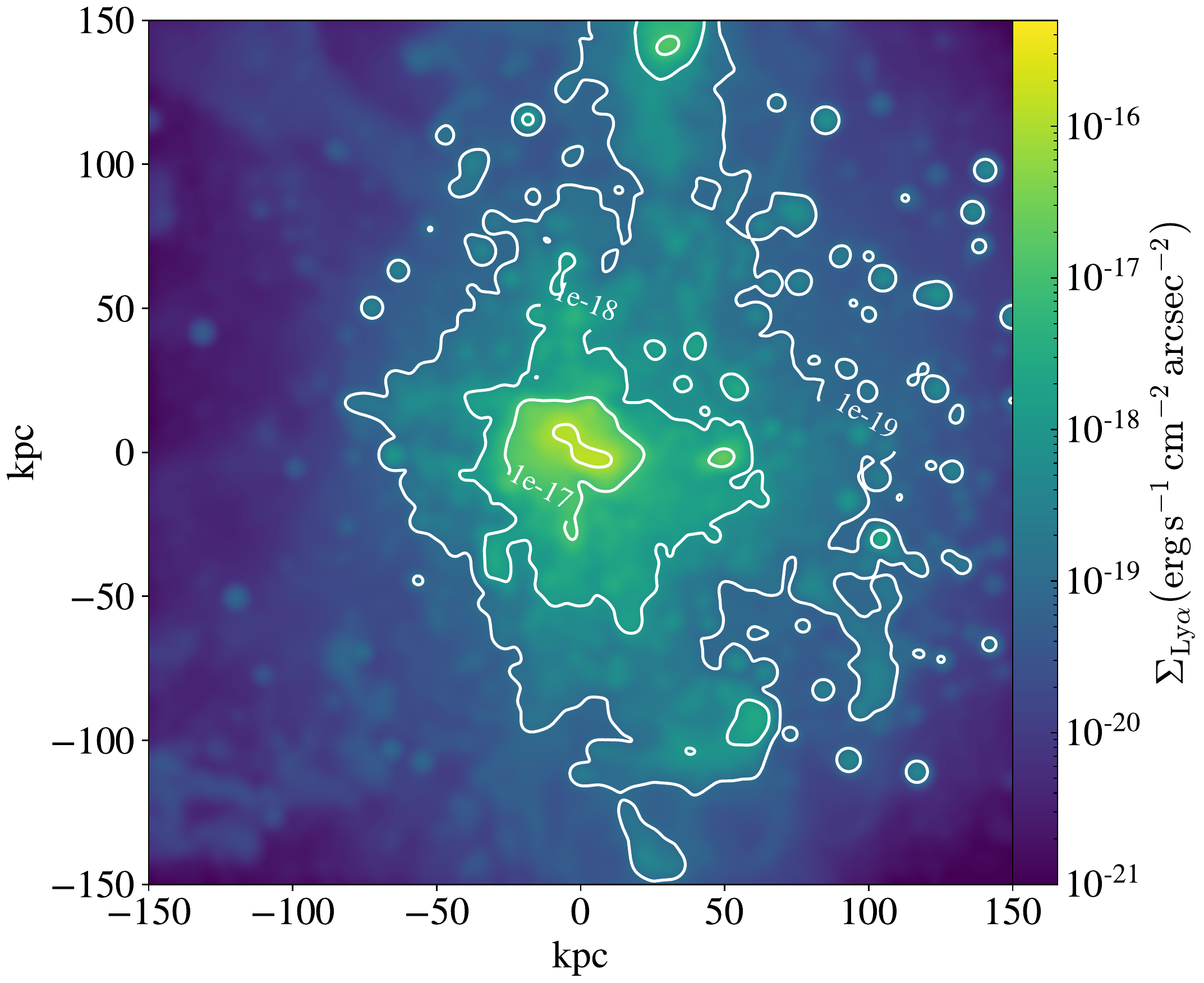}
  \caption{The Ly$\alpha$ surface brightness contours are at $10^{-16}$, $10^{-17}$, $10^{-18}$, and $10^{-19}$ erg/s. Note that though this simulation domain is much larger than those we have discussed previously, it still does not encompass the full extent of a $10^{-19}$ contour. If one were to observe LABs with such sensitivity, one may find them much more extensive than previously published.}
  \label{fig:largebox}
\end{figure}

The size scales of observed LABs have no strict definition, and indeed their highly irregular morphologies makes deriving a simple radius nearly impossible.  Instead, we advocate defining the size as the on-sky area within particular surface brightness contours.  In Figure~\ref{fig:area_plot}, we demonstrated that the areas of blobs both grow with time, i.e. at later times, the enclosed areas are generally larger. This is likely due to the increased availability of scattering gas, possibly as well as the growth of the SFR with time in massive halos. The sizes of these blobs increase with decreasing surface brightness cutoffs. We show this both in Figure~\ref{fig:area_plot}, as well as in Figure~\ref{fig:largebox}, where we compute surface brightness contours for halo A4 down to a limiting surface brightness of $10^{-19}$ erg/s/cm$^2$/arcsec$^2$.

\subsection{On the Origin of the Scattering Gas}
\begin{figure}
  \centering
  \includegraphics[width=\columnwidth]{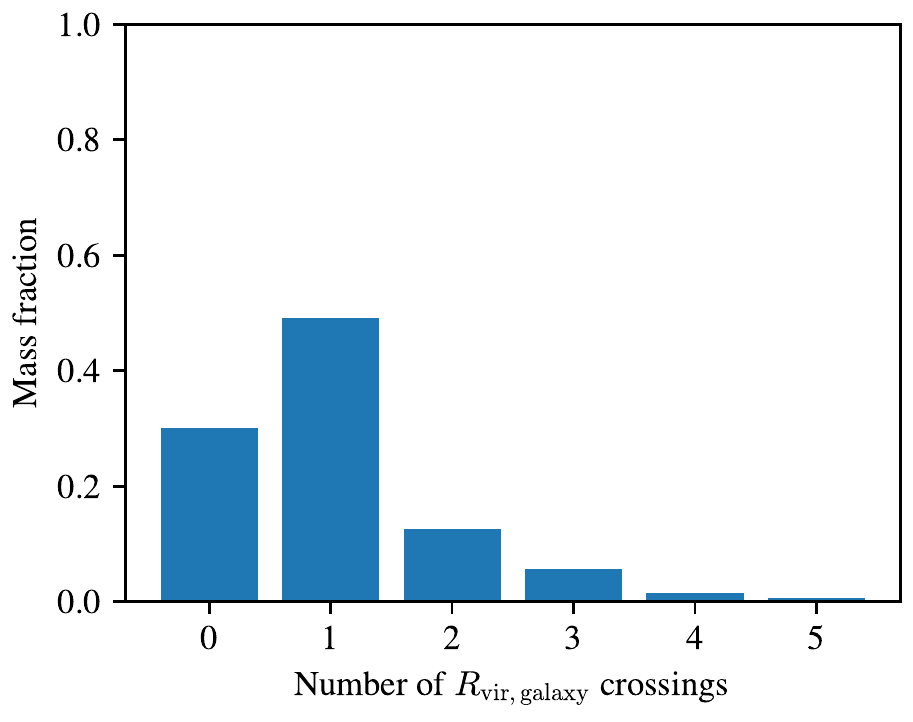}
  \caption{The fraction of gas that is part of the $z=2$ halo but not the $z=2$ central galaxy, broken up by the number of times the gas crosses the $z=2$ central galaxy's virial radius between $z=2$ and $z=5$.}
  \label{fig:galaxy_crossings}
\end{figure}

We now turn our attention to the physical origin of the scattering gas that drives the large scale spatial extents of LABs in our simulations.
Specifically; is the scattering gas pristine gas newly infalling into the halo, or has it been recycled through the central galaxy at some point?
To answer this, we examine our fiducial LAB at $z=2$, and track the number of times that a given gas particle in the CGM (defined in \S~\ref{section:general_physical_properties}) of this galaxy has crossed the central galaxy's virial radius.
We therefore do not distinguish in this case between diffuse CGM, and gas associated with subhalos inside the main halo.
To get some grasp on the historical dynamics of the gas, we track the number of times each CGM particle crosses the central galaxy's virial radius.
In Figure~\ref{fig:galaxy_crossings},  we show the mass fraction of particles as a function of the number of central galaxy $R_{\text{vir}}$ crossings.

The majority of the scattering gas in this example LAB has crossed the virial radius at least once: only $\sim 30\%$ of the gas is pristine, while approximately $50\%$ has passed through the central only once.
While it is difficult to ascertain if the gas that has crossed the virial radius was ejected in a {\it bona fide} outflow (compared to, e.g. either never having been dynamically bound, or simply being at the edge of the friends of friends galaxy finder between snapshots), the relatively large fraction ($\sim 70\%$) of gas that has crossed the virial radius (and therefore been a member of the central galaxy) at some earlier time suggests that the overall baryon cycle between the central galaxy and CGM is important in driving the formation of high-redshift Ly$\alpha$ blobs.

\section{Discussion}
\label{sec:discussion}
\subsection{Uncertainties in the Model}
In our modeling methods, we have made a number of assumptions that could potentially impact our results.  Here, we explore these in turn.

\subsubsection{Impact of the UV Background}

LABs are noted for their spatial extent and one possible explanation for this extent is concentrated emission in dense regions which is scattered through an optically thick medium and thus dispersed across a wide structure. By contrast, it is possible that the spatial extent of a blob may be due to extended emission; namely the outer regions of a gaseous halo that have been ionized, most likely by the cosmological UV background.

To test the impact of our assumed \citet{Faucher-Giguere2009} UV background we conduct a few experiments.  First, we run \textsc{colt} with the ionization state of the gas computed without the presence of any UV background. We find this has no effect on the observable luminosity of the blob, but this effect could be interpreted as evidence that the UV background is unimportant, or that the primary effect of the UV background is the photo-heating effect that is included in the simulations which we cannot remove. Therefore, we concentrate the remainder of our efforts on examining the effect of extremely intense UV backgrounds.

\begin{figure}
  \centering
  \includegraphics[width=\columnwidth]{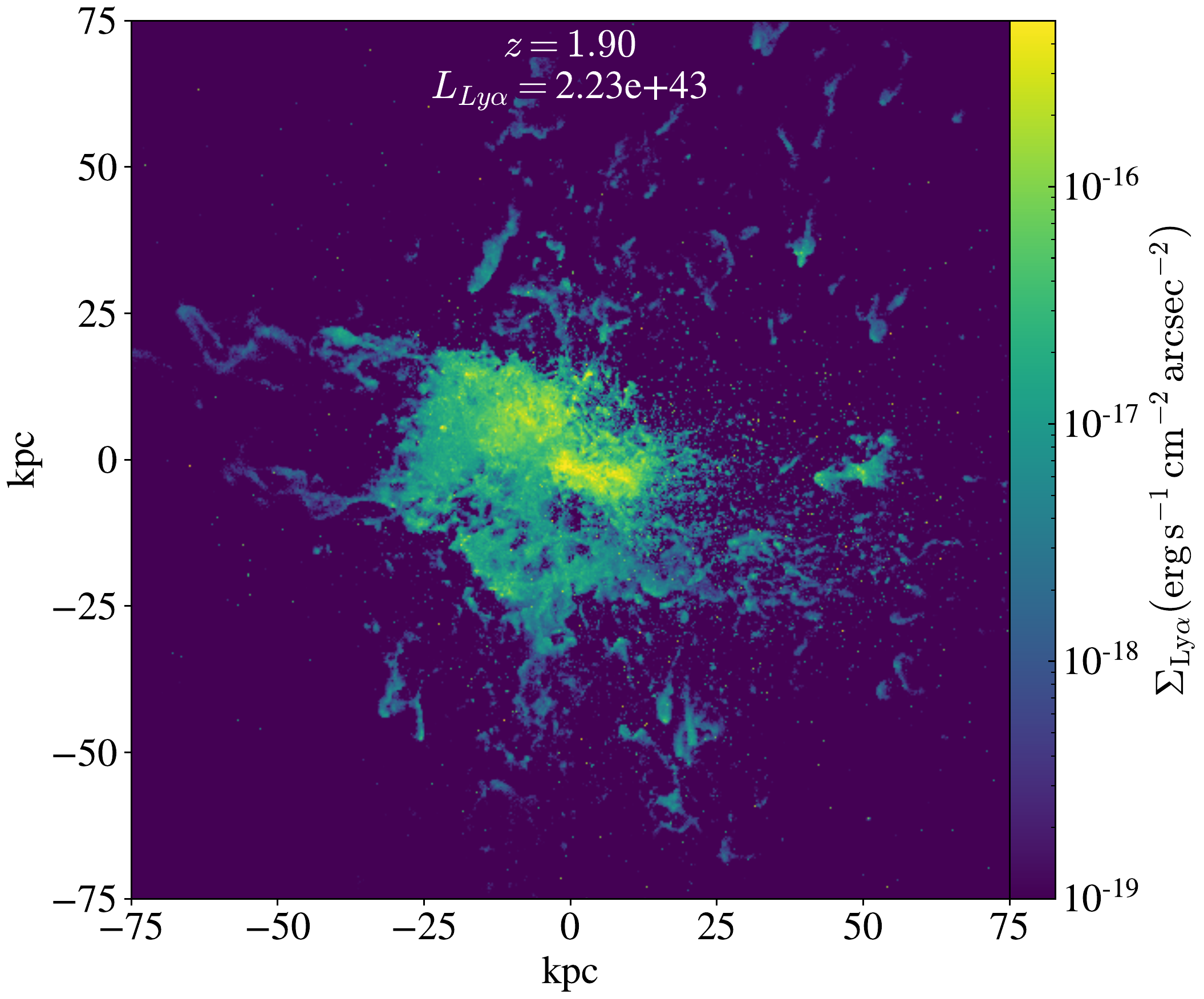}
    \caption{The morphology of a Ly$\alpha$ blob is eroded by using a stronger UV background. The primary effect of additional ionization due to background is to cause the light emitted by star formation activity to escape faster. Compare to Figure~\ref{fig:MUSE}, which uses the \citet{Faucher-Giguere2009} UV background.}
  \label{fig:uvb_morphology}
\end{figure}

In Figure~\ref{fig:uvb_morphology}, we increase the UVB to $1.2\times 10^{-17}\,\text{erg\ cm}^{-2}\text{s}^{-1}\text{Hz}^{-1}$, which is a factor of $10^5$ greater than the fiducial value from \citet{Faucher-Giguere2009} (small enhancements produce no visible effect at all).
At this very elevated background, we see an erosion of the blob due to a reduced optical depth to scattering to Ly$\alpha$ photons in the CGM.
The surface brightness image produced with the enhanced UV background level is both more compact and brighter; it is less like a blob.
Specifically, the area of the LAB at a surface brightness level of $10^{-18}\,\text{erg}\text{s}^{-1}\text{cm}^{-2}\text{arcsec}^{-2}$ at $z=2$ decreases from $83\,\text{arcsec}^{2}$ to $53\,\text{arcsec}^{2}$.
This argues against any substantial contribution from the UV background, and suggests  that LABs are extended not because the source of their emission is itself extended, but because the Ly$\alpha$ emissions is scattered by the CGM.

\subsubsection{Impact of IGM Transmissivity}

\begin{figure}
  \centering
  \includegraphics[width=\columnwidth]{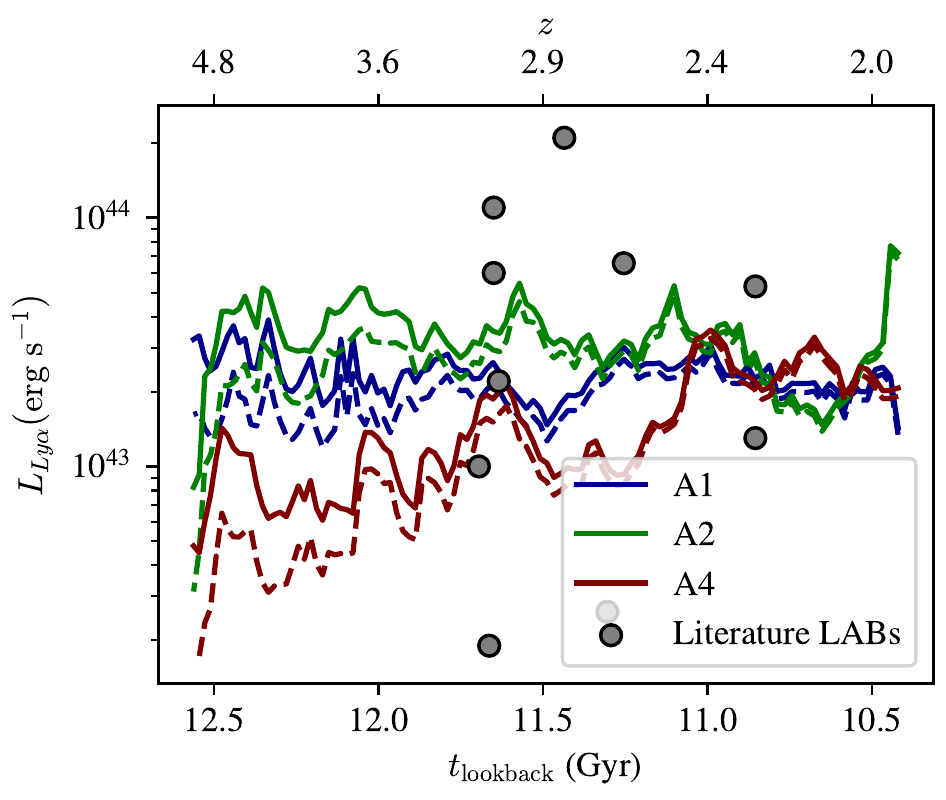}
  \caption{Ly$\alpha$ luminosity for a median line of sight for each galaxy in our sample (lines), with observational data from Table~\ref{table:labs} overplotted as points, and curves for our sample which include IGM extinction in dashed lines.}
  \label{fig:igm_luminosity_redshift}
\end{figure}

In all previous sections we have neglected the effect of IGM extinction.
In this section, we use the frequency-dependent IGM extinction data from \citet{Laursen2011} to compute an IGM transmission fraction for each snapshot based on its escaping spectrum, and plot these in Figure~\ref{fig:igm_luminosity_redshift}.
While the IGM extinction can be substantial, it rarely alters the classification as a LAB.

\subsubsection{Problems with the Hydro Simulations}
We want to call out some specific flaws in the hydro simulations this work is based on that could be improved, and may have an impact on the results presented here.

The UVB photoheating that is used on-the-fly ignores self-shielding. While we are able to recompute the ionization state in post-processing, we cannot back out the temperature increase caused by the lack of self-shielding and then do ionizing radiative transfer more accurately.

While these simulations are zoom-in simulations, they almost certainly still suffer from significant resolution issues. With a higher resolution simulation we should be able to resolve more structure in the CGM, and resolve the so-called multiphase CGM.

These simulations are also missing AGN feedback entirely. In this work we have demonstrated that the presence of an AGN has a large impact on the Ly$\alpha$ properties of LABs, so it is very likely that the thermal feedback, kinematics, and impact on star formation produced by AGN are relevant too.

\subsection{Comparison to Other Theories of LAB Formation}
Our model includes the physics of emission from ionized gas surrounding star-forming regions, cooling from gas excited by collisions, and recombinations from ionized gas.
As discussed in \S\ref{sec:agn}, we additionally (optionally) include the potential impact of AGN on the Ly$\alpha$ luminosity from our model halos.
This model builds on a rich history of literature models in this field that typically include only a subset of the aforementioned physics.
In what follows, we discuss the findings of these models grouped by the physics they include.

Some of the previous work on LAB formation has been focused strongly on emission from cooling gas, that is, collisionally excited neutral hydrogen.
\citet{Fardal2001} simulate only cooling emission and find luminosities of about the correct order of magnitude to be blobs, but would need to invoke yet-unquantified radiative transfer effects to explain the size.
\citet{Haiman2000} find that cooling emission is sufficient to reproduce the blobs mentioned in \citet{Steidel2000} (but their model is a simple analytical one).
\citet{Faucher-Giguere2010} also limit their model to cooling emission, but note that to reproduce the luminosities observed they need to model cooling emission from regions which should be experiencing star formation.
They do perform radiative transfer, and are able to reproduce the spatial extent of LAHs and the characteristic line profile of Ly$\alpha$ nebulae.
\citet{Rosdahl2012} study emission from cooling gas but additionally incorporate recombinations by treating them as a source of cooling, but are thus unable to discuss luminosity driven by recombinations.

In contrast to these works, \citet{Cantalupo2005} and \citet{Gronke2017} omit cooling emission from their model, but include emission from recombinations.  We note that in our own model, for the non-AGN case, cooling emission dominates the Ly$\alpha$ luminosity.
\citet{Kollmeier2010} model emission due to the cosmological UV background and the presence of a quasar (which they term fluorescence).

Some papers in the literature account for both emission from cooling and recombinations \citep{Cen2013, Furlanetto2005, Geach2016, Smailagi2016}.
However, in all these papers the Ly$\alpha$ luminosity from recombinations is determined only by star formation rates. 
An exception to this is the work of \citet{Geach2016}, which accounts for the contribution of ionization from stars by locally modifying the ionization state of the gas based on the local star formation rate.
In this work, we use stellar population synthesis on star particles from the simulations we are post-processing to compute an ionizing radiation field, and thus the ionization state of our gas.
This different approach makes it possible to capture RT effects from the ionization state calculation as well.

Some of the previous attempts at LAB modeling do not include Ly$\alpha$ radiative transfer effects \citep{Fardal2001, Furlanetto2005, Goerdt2010, Smailagi2016, Rosdahl2012}.
For the existing work that does include RT, it is often restricted due to low spatial resolution \citep{Cen2013} or discuss only a single line of sight \citep{Cantalupo2005}.
In this paper we demonstrate that there are strong line-of-sight variations, which can only be reproduced with reasonably accurate RT calculations.

We now turn to whether other theoretical works for LAB formation agree with our own findings. Generally, all of the aforementioned studies that aim to model LABs  find blob-like objects, but a few have notable results which are in agreement with our findings.
\citet{Laursen2007} find large surface brightness variation due to escape anisotropy in a simulated LBG.
\citet{Cen2013} finds ubiquitious blob-like objects around massive halos, as do we (albiet with a relatively small sample size).
\citet{Kollmeier2010} report cosmological-scale Ly$\alpha$ at a surface brightness of $10^{-19}$ erg/s/cm$^2$/arcsec$^2$, due to fluorescence.
This result is quite similar to our own findings.

\section{Conclusions}
\label{sec:conclusions}
We have combined high-resolution cosmological zoom simulations of massive galaxy evolution at high-redshift with $3$D Monte Carlo Lyman-$\alpha$ radiative transfer simulations to develop a model for the origin of Lyman-$\alpha$ blobs at high-redshift ($z = 2$--$5$).
Our work considers the physics of ionization radiative transfer, cooling emission, recombination, and the impact of AGN within the cosmological context of galaxy evolution.
Our main conclusions from this work are as follows:
\begin{itemize}

    \item When adopting a notional luminosity-based definition for LAB formation, we find that all of our model massive galaxies go through a LAB phase at some point between $z = 2$--$5$ (see Figures~\ref{fig:luminosity_redshift}, \ref{fig:agn_comparison}, \ref{fig:MUSE}, and \ref{fig:grid_plot}).  These LABs have extended morphologies in agreement with observations.

    \item The escape fraction of Lyman-$\alpha$ photons from these objects are highly orientation-dependent, which complicate our observational understanding of the connection between Lyman-$\alpha$ blobs and massive galaxy evolution. Variations in the escape fraction with respect to different lines of sight can produce
    large variations in the observed Ly$\alpha$ luminosity (see Figure~\ref{fig:los_variation}).
    
    \item The formation of LABs in our model is independent of the inclusion of AGN: star formation alone is enough to drive LAB formation in massive galaxies. This said, the presence of an AGN can significantly enhance Lyman-$\alpha$ luminosities and alter spatial extents (see Figures~\ref{fig:agn_comparison} and \ref{fig:area_plot}).
    
    \item The presence of an AGN in a LAB may be detectable by measuring the spatial concentration of Ly$\alpha$ luminosity (see Figures~\ref{fig:area_plot} and \ref{fig:skewness}).
    
    \item The observed LAB luminosities do not scale very strongly with any individual physical property except for stellar mass.  The reason is that the intrinsic luminosity is dependent on the temperature and ionization state of the gas, which can vary strongly with the small scale geometry of the gas distribution.  Similarly, the observed luminosity folds in the escape fraction which is a strong function of the star-gas-dust geometry, as well as the small scale clumping  (see Figure~\ref{fig:luminosity_vs_temperature}).
    
\end{itemize}

Future improvements in the model could include both a full radiative hydrodynamic treatment of the galaxy formation simulations, as well as \emph{bona fide} AGN feedback.
Similarly, future generations of models that include lower mass halos may be able to connect Ly$\alpha$ blobs, Ly$\alpha$ emitters, and Ly$\alpha$ halos.

\section*{Acknowledgments}
We thank Claude-Andr\'{e} Faucher-Gigu\`{e}re and George Privon for helpful conversations.
DN acknowledges support from grants HST-AR-13906.004-A, HST-AR-15043.001-A from the Space Telescope Science Institute, and NSF AST-1715206.
AS acknowledges support for Program number HST-HF2-51421.001-A provided by NASA through a grant from the Space Telescope Science Institute, which is operated by the Association of Universities for Research in Astronomy, Incorporated, under NASA contract NAS5-26555.
RF acknowledges financial support from the Swiss National Science Foundation (grant no 157591).
The simulations were run using XSEDE (TG-AST160048), supported by NSF grant ACI-1053575, Northwestern University's compute cluster `Quest', and on the University of Florida HiPerGator computing cluster.
The data used in this work were, in part, hosted on facilities supported by the Scientific Computing Core at the Flatiron Institute, a division of the Simons Foundation.
This work was initiated or performed in part at the Aspen Center for Physics, which is supported by National Science Foundation grant PHY-1607611.


\bibliographystyle{mnras}
\bibliography{biblio}


\appendix

\section{Impact of temperature on our model results}
\label{app:half_temperature}

\begin{figure*}
    \centering
    \includegraphics[width=\textwidth,height=\textheight,keepaspectratio]{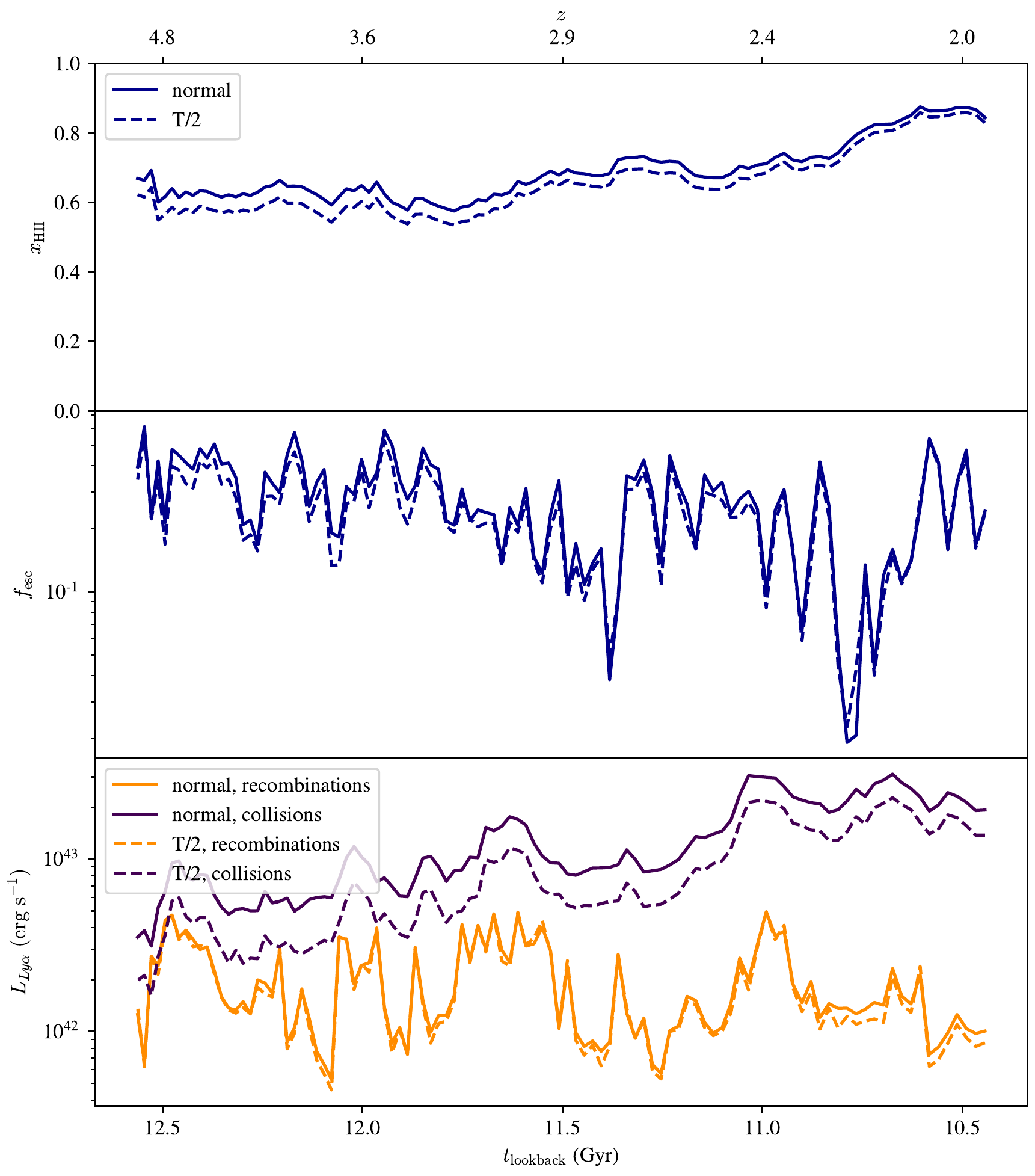}
    \caption{
        To assess the impact of the lack of the on-the-fly self-shielding approximation on the Ly$\alpha$ luminosity of our gas, we manually decrease the temperature  of all gas in the simulation by a factor of 2, before computing the ionization state and doing Ly$\alpha$ radiative transfer. The luminosity of our fiducial blob is slightly decreased.}
    \label{fig:half_temperature}
\end{figure*}

Since the gas in the snapshots we are processing is potentially overheated due to the crude model for self-shielding in the hydrodynamic simulations, we examine the impact of halving the gas temperature in Figure~\ref{fig:half_temperature} before the ionization state calculation or Ly$\alpha$ radiative transfer.
We find that the luminosity due to collisions is decreased by about a factor of 1.5.
There are at least two factors at play here: The effect on ionization state and the effect on the gas temperature relation shown in Figure~\ref{fig:luminosity_vs_temperature}.
The decrease in ionization state results in a decrease in the availability of free electrons to collisionally excite neutral hydrogen.
Additionally, the decreased ionization state (mostly) decreases the escape fraction, which overall decreases the luminosity of the blob.
In the bottom panel of Figure~\ref{fig:half_temperature} we can see that the impact on emission due to collisional excitations is much more substantial than the effect of the impact on recombinations, but that a factor of 2 change in temperature is not sufficient to invalidate any of the previous conclusions in this work.
That is, emission due to collisions dominate, and this has a luminosity typical of LABs.

\section{Physical Origin of Escaping Ly\texorpdfstring{$\alpha$}{a}}
\label{app:correlations}

\begin{figure*}
    \centering
    \includegraphics[width=\textwidth,height=\textheight,keepaspectratio]{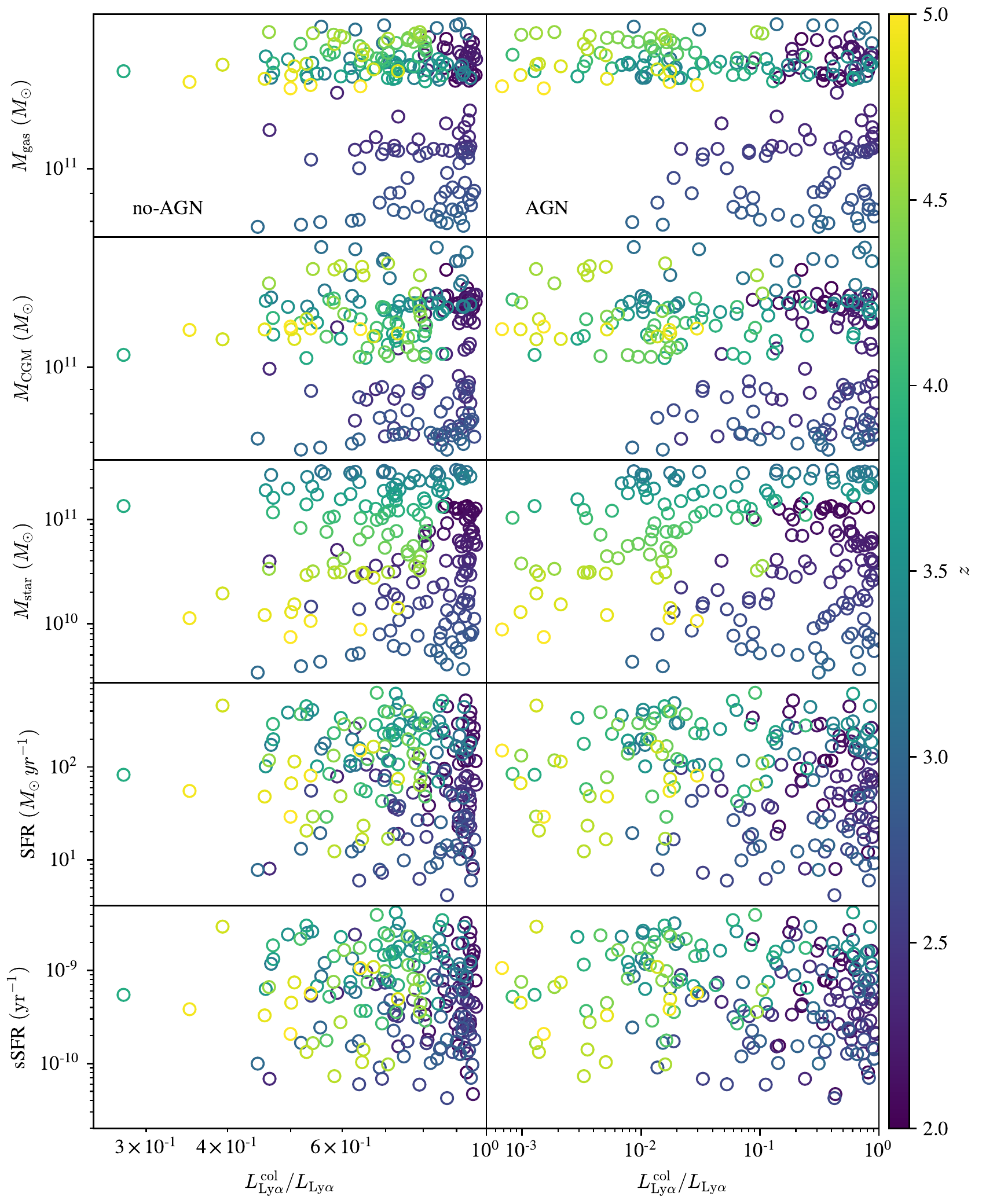}
    \caption{
        We plot the fraction of escaping Ly$\alpha$ due to emission from collisionally excited hydrogen and recombinations as a function of various physical properties of the galaxy that one might expect to predict which source dominates the luminosity.
        The left column is for the mode where we do not account for AGN, and on the right we account for ionization due to AGN.
        Note the difference in the horizontal axis limits.
    }
    \label{fig:appendix_correlations}
\end{figure*}

In \S~\ref{sec:emission_in_simulations} we noted that there are no strong trends between halo physical properties and whether emission from recombinations or emission from collisional excitations dominate, and here we include Figure~\ref{fig:appendix_correlations} to illustrate that explicitly.

As we have discussed prevoiusly in this work, the {\it escaping} Ly$\alpha$ luminosity, and variations in this escape fraction can be sufficient to confound simplistic reasoning about the powering source of blobs.
Increasing the intrinsic emission of Ly$\alpha$ is not sufficient to drive observed (that is, escaping) luminosity of the halo if the escape fraction of that increased luminosity decreases or is stochastic.
For example, a growing population of neutral hydrogen will augment the collisional component of Ly$\alpha$, but with that increased emission comes more Ly$\alpha$ optical depth, and thus more potential for the emitted Ly$\alpha$ to be absorbed by dust, depending on the geometry of the system.
Further complicating this issue of physical property correlations is that the escape fraction of emission from collisionally excited hydrogen and recombinations differ, often by more than a factor of 2.

\section{Testing the impact of our assumed core-skipping approximation}
Throughout this work, we have implemented a core-skipping algorithm for efficiency purposes (it improves runtime by approximately an order of magnitude).  Core-skipping is an approximation wherein photon packets which are in the core of the Ly$\alpha$ line and also in a cell which is optically thick along all lines of sight have their frequency shifted farther from line center.
Here, we run a single snapshot without core skipping to get a concrete idea of its effects on our results.  In this test, we see that turning off core skipping results in a small increase in the median escape fraction of about 3\%.

\label{lastpage}
\end{document}